\documentclass[twocolumn,footinbib,prx,superscriptaddress]{revtex4-2} %
\usepackage{xcolor}
\usepackage{amsfonts,amssymb,amsmath,mathrsfs}
\usepackage{comment}
\usepackage{graphicx}
\usepackage[colorlinks]{hyperref}%
\hypersetup{%
    plainpages=true,
    breaklinks=true,
    hypertexnames=false,
    pageanchor=true,
    colorlinks=true,
    linkcolor={blue},
    citecolor={cyan},
    urlcolor={cyan},
    anchorcolor={black}
}

\newcommand{\xsection}[1]{\section{#1}}

\usepackage[utf8]{inputenc}

\allowdisplaybreaks

\begin{document}

\title{Quantum dynamics of semiconductor quantum dot Josephson junctions}

\author{Utkan Güngördü}
\email{utkan@lps.umd.edu}
\author{Rusko Ruskov}
\affiliation{Laboratory for Physical Sciences, 8050 Greenmead Drive, College Park, Maryland 20895}
\affiliation{Department of Physics, University of Maryland, College Park, Maryland 20742, USA}
\author{Silas Hoffman}
\affiliation{Laboratory for Physical Sciences, 8050 Greenmead Drive, College Park, Maryland 20895}
\author{Kyle Serniak}
\affiliation{Lincoln Laboratory, Massachusetts Institute of Technology, Lexington, MA 02421, USA}
\affiliation{Research Laboratory of Electronics, Massachusetts Institute of Technology, Cambridge, MA 02139, USA}
\author{Andrew J. Kerman}
\affiliation{Lincoln Laboratory, Massachusetts Institute of Technology, Lexington, MA 02421, USA}
\author{Charles Tahan}
\altaffiliation[Current address: ]{Department of Physics, University of Maryland, College Park, Maryland 20742, USA}
\affiliation{Laboratory for Physical Sciences, 8050 Greenmead Drive, College Park, Maryland 20895}

\begin{abstract}
Josephson junctions constructed from superconductor-semiconductor-superconductor
heterostructures have been used to realize a variety of voltage-tunable superconducting quantum
devices, including qubits and parametric amplifiers.
To date theoretical descriptions of these systems have been restricted to
small quantum fluctuations of the junction phase,
making them inapplicable to   %
many experiments.
In this work we relax this,  %
employing a path-integral formulation
where the phase quantum dynamics   %
is obtained self-consistently
from an underlying many-body formalism.
Our method recovers previously-known results    %
for
small phase fluctuations,
and predicts   %
new effects outside of that limit:
(i) system capacitances  %
undergo
a gate-voltage-dependent renormalization;
and (ii) an additional   %
charge offset appears for asymmetric junctions.
Our main results can be
summarized in terms of   %
a
single-particle Hamiltonian,
which
can be directly compared
to that of an ordinary Josephson junction.
This more general theory could be   %
a
first step towards
designing new quantum devices that go qualitatively beyond voltage-tunable variants of previously-known circuits.
\end{abstract}
\maketitle

\xsection{Introduction}
Superconducting circuits based on Josephson junctions (JJs) have been used to realize a variety of devices,
including quantum amplifiers, digital logic circuits, and photon detectors.
The most intensively studied applications of these circuits, however, are qubits,
making these one of the    %
promising and fastest-growing platforms for realizing large-scale
quantum information processors~\cite{kjaergaard_superconducting_2020,bravyi_future_2022}.
Many different JJ-based qubits have been demonstrated, including the transmon~\cite{koch_charge-insensitive_2007},
flux qubit \cite{doi:10.1126/science.285.5430.1036},
fluxonium~\cite{manucharyan_fluxonium_2009}, and
$0-\pi$ qubit~\cite{gyenis_experimental_2021}.
The JJs    %
of these circuits have most often been based on
superconductor-insulator-superconductor (S-I-S)
tunnel-junctions; %
however, high-quality superconductor-semiconductor  %
(super-semi) heterostucture JJs
have also recently been realized in
a
variety of
materials.
These super-semi junctions, while exhibiting
the Josephson effect,
carry
critical currents that are tunable via the field effect of an electrostatic gate,
allowing a new class of voltage-tunable  quantum  circuits such as the
gatemon~\cite{larsen_semiconductor-nanowire-based_2015,de_lange_realization_2015, pita-vidal_gate-tunable_2020},
Andreev-pair qubits~\cite{janvier_coherent_2015,hays_direct_2018} and
Andreev spin qubits~\cite{hays_coherent_2021-1,pita-vidal_direct_2023}.
These devices have also served as a testbed for the underlying Andreev physics of superconducting weak links,
and more recently, as building blocks of new topologically
non-trivial superconducting circuits \cite{matute2023quantum,PinoSoutoAguado2024}.

Much of the theory of super-semi junctions has built upon the Bogoliubov--de Gennes picture of Andreev transport
in superconducting-normal-superconducting junctions~\cite{beenakker_josephson_1991},
with extensions for the practically relevant situation in which an electrostatic disorder potential forms
a quantum dot in the junction.
The low energy dynamics of these S-QD-S junctions have been studied
in circuits
where
the junction is shunted by a small inductance,
wherein the gauge-invariant phase difference between superconducting electrodes is set,
apart from small quantum fluctuations,
by the external flux through the inductance \cite{zazunov_andreev_2003, kurilovich_microwave_2021}.
This regime is relevant for Andreev qubits.
However, these theories do not extend to S-QD-S junctions embedded in arbitrary circuit environments
that may support strong
quantum phase fluctuations across the junction (e.g. transmon or fluxonium circuits).

In this paper, we develop a microscopic theory of
a S-QD-S junction
embedded in
a capacitive environment in which quantum fluctuations of the phase may be large.
Our self-consistent treatment of quantum dynamics of the superconducting phases reveal two effects
that originate from the underlying many-body physics: (i) a renormalization of the
effective
capacitance that shunts the junction and (ii)
appearance
of an additional charge offset
in the charging energy
for asymmetric junctions.
The dependence of both effects on junction gate voltage make them
important
for
the
analysis, control, and design of superconducting circuits with S-QD-S junctions.

\xsection{Many-body treatment of superconducting circuits} In the phenomenology of JJs and circuit quantization
\cite{Likharev-book-1986,Devoret1997-Les_Houches},
JJs are treated as non-linear inductors,
and quantization is postulated from the corresponding
classical Hamiltonian of the circuit \cite{Devoret1997-Les_Houches}.
Within this framework, the well-known Hamiltonian for a capacitively shunted JJ
(such as a Cooper pair box (CPB) or transmon \cite{koch_charge-insensitive_2007}) is obtained as
\begin{align}
\hat H_\text{CPB} = \frac{(2e)^2}{2 C_\Sigma}\left[\hat{n} - n_g(V^a) \right]^2 - E_J \cos (\hat\phi),
\label{CPB-Hamiltonian}
\end{align}
reflecting the quantum mechanics
of the phase difference
between
the two superconducting leads $\phi \equiv \phi_L - \phi_R$,
where the phase $\phi$ plays the role of ``coordinate'' for the Josephson potential, and the charging energy
looks like a classical expression
with shunting capacitance $C_\Sigma$,
a
dimensionless
charge operator, $\hat{n} \equiv -i \partial_\phi$
and a charge offset $n_g(V^a)$
associated with an externally applied voltage $V^a$.

\begin{figure}
    \centering
    \includegraphics[width=0.95\columnwidth]{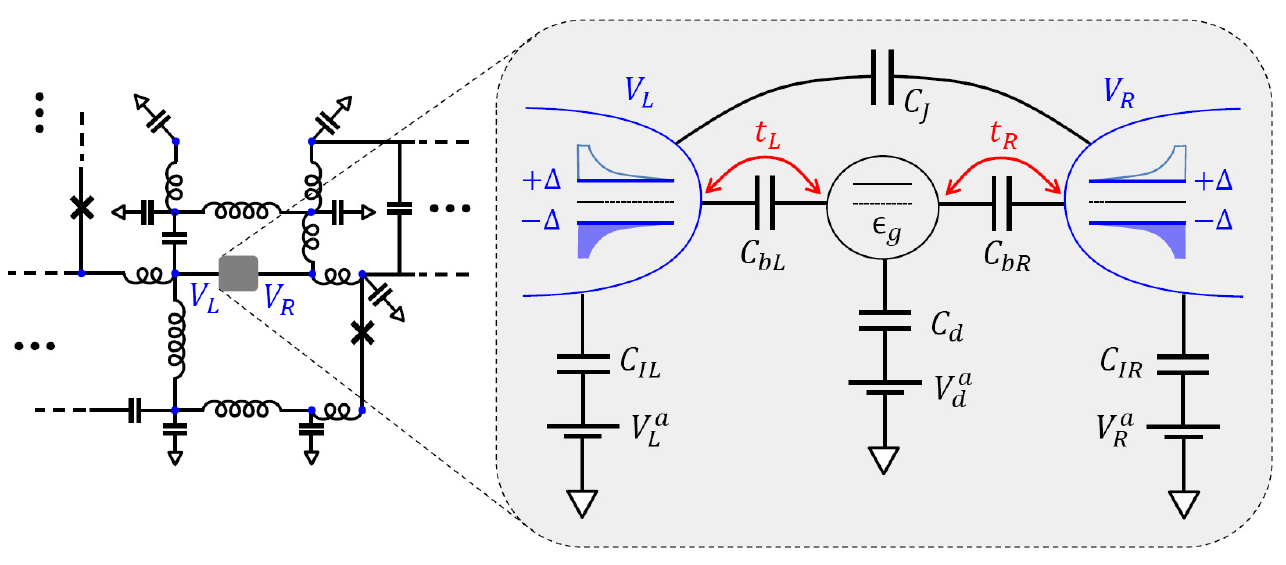}
    \caption{A capacitively shunted S-QD-S junction embedded in a general circuit environment
    which does not restrict phase dynamics. The leads (blue, with BCS density of states)
    are tunnel coupled
    via $t_L,t_R$
        to the QD (containing a single level).
    $C_J$ is the capacitance between the leads, and $C_{bL}$,$C_{bL}$ are capacitances between
    the dot and each lead. $V_L$ and $V_R$ denote the mean field voltage variables for each lead,
    and $\epsilon_g$ is the effective energy of the dot level, derived in the text.
    $V_L^a, V_R^a, V_d^a$ are externally applied voltages across capacitances $C_{IL}$, $C_{IR}$, $C_d$ respectively.
    For simplicity, we consider a symmetric
        capacitance arrangement:
    $C_{bL}=C_{bR}=C_{b}$, $C_{IL} = C_{IR} = C_I$.}
    \label{fig:circuit}
\end{figure}

This Hamiltonian, $\hat H_\text{CPB}$, can be derived self-consistently from an underlying many-body theory
where a pair of tunnel-coupled superconducting leads are described by BCS Hamiltonians,
and
charging energy is introduced in a many-body picture.
This program was realized in a seminal paper
by Ambegaokar, Eckern, and Sch\"{o}n
\cite{ambegaokar_quantum_1982},
where the path integral formulation of a (grand canonical) partition function
$Z_G \propto \text{tr} e^{-\beta \hat H }$
($\beta =1/k_B T$ is the inverse temperature)
is evaluated at a saddle point, generating (self-consistently) the superconducting order parameters of the leads,
$\Delta e^{i \phi_{L,R}}$,
and the voltage drop across the junction, $V = V_L - V_R$, both in a mean field approach.
The partition function is then reduced to an effective action,
$Z_G \sim \int \mathcal D\phi e^{ - \frac{S[\phi(\tau)]}{\hbar} }$,
from which the Hamiltonian $\hat H_\text{CPB}$ can be deduced under the ``slow phase approximation"
in which $\phi(\tau)$ varies slowly over the time scale $\tau \sim \hbar/\Delta$.
In the next order of slow phase expansion, Eckern et al.
\cite{eckern_quantum_1984},
and Larkin and Ovchinnikov \cite{larkin_decay_1983},
derived a small renormalization of the shunting capacitance
across the junction, $\delta C_\Sigma^\text{JJ} = 3\pi \hbar/(32 \Delta R_N)$
where $R_N$ is the normal state resistance of the junction
\cite{eckern_quantum_1984}.
This
capacitance does not significantly alter the quantization prescription for
such circuits \cite{Devoret1997-Les_Houches};
at the same time
its smallness makes it difficult to measure experimentally.

To summarize the results of this paper, we use the formalism of
Ref. \cite{ambegaokar_quantum_1982,eckern_quantum_1984} to describe a S-QD-S junction in a
capacitive environment (inset Fig.~\ref{fig:circuit}) and obtain an effective Hamiltonian with a form
that is similar to $\hat H_\text{CPB}$,
given by
\begin{eqnarray}
&&\hat H_\text{even} =
\frac{(2e)^2}{2 (C_{\Sigma} + \delta C_{\Sigma} ) }\left[\hat{n}
- \hat{n}_q(V^a,\Gamma_{L,R},\epsilon_g,\Delta) \right]^2
\nonumber\\
&& \qquad\quad { } + \hat{U}_J(\phi,\Gamma_{L,R},\epsilon_g,\Delta) .
\label{new supersemi_Hamiltonian}
\end{eqnarray}
Here, the Josephson potential, $\hat U_J$, and
the charging Hamiltonian are matrices acting on the even occupancy (singlet) space of the dot,
$\{|0\rangle,|\uparrow\downarrow\rangle\}$, and depend on the dot's gate voltage, $\epsilon_g$ and
the tunneling rates $\Gamma_{L,R}$ between the leads and the dot.
The \emph{derived} charge offset ($\hat n_q$) differs from the \emph{assumed} form ($n_g$) in the
literature
\cite{kringhoj_suppressed_2020-1,bargerbos_observation_2020-1,vakhtel_quantum_2023}:
it contains new terms that depend on tunneling asymmetry and dot occupation
(proportional to the Pauli matrices $\eta_0$ and $\eta_z$),
moreover, the capacitance renormalization $\delta C_{\Sigma}(\Gamma_{L,R},\epsilon_g,\Delta)$ is
gate-voltage
tunable.
The Josephson potential matrix $\hat U_J$, whose eigenvalues determine the Andreev bound states (ABS) spectrum,
essentially coincides with the results of \cite{kurilovich_microwave_2021},
with the addition of
non-perturbative corrections for finite dot voltage $\epsilon_g$.

\xsection{Model}
By quantizing the classical Hamiltonian of the circuit shown in the inset of Fig.~\ref{fig:circuit}
and combining it with the junction Hamiltonian, we obtain $\hat H = \hat H_J + \hat H_Q$, where
\begin{widetext}
\begin{align}
\hat H_J =& \sum_{i=L,R,\sigma=\uparrow,\downarrow}  \int dr \left( \hat \psi_{i,\sigma}^\dagger \hat \xi_i \hat\psi_{i,\sigma}
 -\frac{g}{2} \hat\psi_{i,\sigma}^\dagger \hat\psi_{i,-\sigma}^\dagger \hat\psi_{i,-\sigma} \hat\psi_{i,\sigma} \right) + \hat d_\sigma^\dagger \mu_d \hat d_\sigma + \left(t_i   \hat d_\sigma^\dagger  \hat \psi_{i,\sigma}(r=0) + \text{H.c}\right), \nonumber\\
 \hat H_Q =& \frac{1}{2 C_\Sigma} \left(\frac{\hat Q_L-\hat Q_R}{2}\right)^2
    -\frac{1}{C_\Sigma} \left(\frac{\hat Q_L- \hat Q_R}{2}\right) \Delta Q +   \epsilon_d \frac{1}{e} \hat Q_d + \frac{1}{e^2} U \hat Q_d^2. %
\end{align}
\end{widetext}
$\hat H_J$ models the junction (dot and leads) and
$\hat H_Q$
describes a capacitive  circuit environment
in terms of the charges
$\hat Q_i = \sum_\sigma e  \int dr \hat \psi_{i,\sigma}^\dagger   \hat \psi_{i,\sigma}$ and
$\hat Q_d = \sum_\sigma e \hat d_\sigma^\dagger \hat d_\sigma$.
The fermionic field operators for the leads and dot are respectively $\hat \psi_{i,\sigma} = \hat \psi_{i,\sigma}(r)$
and $\hat d_{\sigma}$ with spin $\sigma$, where the dot is modeled as having a single accessible level
\footnote{Assuming the
dot's energy quantization provides the largest energy scale,
$\delta E \gg \Delta, U$, scf. \cite{kurilovich_microwave_2021} }.
Above, $\hat\xi_i = \frac{\hat p^2}{2 m^*} - \mu_i$ is the kinetic energy operator for the leads
with effective mass $m^*$, $\mu_i$ and $\mu_d$ are chemical potentials for isolated leads and dot,
$g$ is the strength of the
BCS
pair potential around the Fermi level,
and
$t_i$ is the tunneling strength between the leads and the dot.
The capacitance across the junction, $C_\Sigma = C_J + \frac{C_b + C_I}{2}$,
the charging energy of the dot $U = \frac{e^2}{4 \left(C_b + \frac{C_d C_I}{C_d + 2C_I} \right)}$,
the charge offset, $\Delta Q = \frac{C_I}{2} V^a$ due to
an
applied voltage  $V^a = V_R^a - V_L^a$, and
the shift in the dot level
$\epsilon_d = \frac{4 U}{e} \frac{C_I C_d}{2 C_I + C_d} \left(V_d^a - \frac{V_R^a + V_L^a}{2}\right)$,
are  due to the electrostatic environment.%

To capture the quantum dynamics of the phase
difference between the electrodes,
we express the partition function of the system,
$Z_G = \text{tr}e^{-\beta \hat H}$, as an imaginary-time %
($\tau$)
fermionic coherent state path integral
\cite{ambegaokar_quantum_1982,larkin_decay_1983,zazunov_andreev_2003,coleman_introduction_2015,altland_condensed_2010}.
We eliminate all quartic interaction terms using the Hubbard--Stratonovich transformation at the expense of
introducing auxiliary bosonic fields $\Delta(\tau) e^{i \phi_i(\tau)}$, $V_i(\tau)$, and $M(\tau)$,
representing the $s$-wave superconducting order
parameters \cite{ambegaokar_quantum_1982,eckern_quantum_1984,coleman_introduction_2015},
the
voltage of the leads, and the magnetic Weiss field \cite{altland_condensed_2010,rozhkov_josephson_1999},
respectively,
followed by saddle point approximations that pin $\Delta(\tau) \to \Delta$,
$i\hbar \partial_\tau \phi_i(\tau) \to 2 e V_i(\tau) $ (i.e., Josephson relation), and $M(\tau) \to M$.
From self-consistent calculations \cite{rozhkov_josephson_1999},
we find the saddle point value of $M$ approximately vanishes for even occupancy states of the dot
at any phase and away from the Kondo regime,
i.e.
for
$\Delta \gg T_K
= \sqrt{ \frac{U \Gamma}{2} } e^{-\frac{\pi}{8 U \Gamma}|U^2 - 4 \epsilon_g^2|  }$
where $\Gamma \equiv \Gamma_L + \Gamma_R$
\cite{meden_andersonjosephson_2019}.
Henceforth, unless otherwise noted, we restrict our analysis to the even occupancy sector
and to a set of parameters which are outside the Kondo regime. Consequently, we take $M=0$,
so the overall effect of the Coulomb interaction is a shift of
the dot level \cite{oriekhov_voltage_2021} by $\frac{U}{2}$,
which is contained in the definition of $\epsilon_g = \epsilon_d + \mu_d + \frac{U}{2}$.

Performing a unitary transformation that shifts all time dependence due to $\phi_i(\tau)$
onto the tunneling terms from the leads \cite{eckern_quantum_1984}, we obtain
\begin{align}
\label{eq:Ginv}
& Z_G =  \int \mathcal D \phi_L \mathcal D \phi_R \mathcal D^2\Psi e^{- \frac{1}{\hbar}
\int_0^{\hbar\beta} d\tau  \sum_{\boldsymbol k} \bar\Psi(\boldsymbol k,\tau) [-G^{-1}(\boldsymbol k,\tau)] \Psi(\boldsymbol k,\tau) }   \nonumber \\
& \times e^{\frac{1}{\hbar}\int_0^{\hbar\beta} d\tau \frac{C_\Sigma}{2} \left(V(\tau)
+ \frac{C_I}{2 C_\Sigma} V^a \right)^2} e^{-\beta \frac{M^2}{2 U} }, \text{ with} \\
& -G^{-1}(\boldsymbol k,\tau) = \hbar\partial_\tau +
\nonumber\\
&\begin{pmatrix}
 \xi_{L,\boldsymbol k} \tau_z + \Delta \tau_x  & 0 & \frac{t_L}{\sqrt{\mathcal V_L}} \tau_z e^{-i \tau_z \frac{\phi_L(\tau)}{2} } \\
0 & \xi_{R,\boldsymbol k} \tau_z + \Delta \tau_x & \frac{t_R}{\sqrt{\mathcal V_R}} \tau_z e^{-i \tau_z \frac{\phi_R(\tau)}{2}} \\
\frac{t_L}{\sqrt{\mathcal V_L}} \tau_z e^{i \tau_z \frac{\phi_L(\tau)}{2}} & \frac{t_R}{\sqrt{\mathcal V_R}} \tau_z e^{i \tau_z \frac{\phi_R(\tau)}{2}} & \epsilon_g  \tau_z + M
\end{pmatrix} \nonumber
\end{align}
where $G(\boldsymbol k,\tau)$ is the Green's function of the junction in
a
momentum representation, the term $\propto C_\Sigma$ is the capacitive energy
with $V(\tau) = V_L(\tau) - V_R(\tau) = \frac{\hbar}{2 e} i \partial_\tau \phi(\tau)$,
and $\Psi(\boldsymbol k,\tau) = \Psi = (\psi_L, \psi_R, D)^T$,
where
$\psi_i = (\psi_{i,\uparrow}, \bar\psi_{i,\downarrow})^T$, $D = (d_\uparrow, \bar d_\downarrow)^T$
are Grassmann--Nambu spinors, and $\mathcal V_i$ is the volume of each lead.
We proceed to obtain a description of the ABS, which have contributions from in-gap and continuum energies.

\xsection{In-gap contributions}
By integrating out the fermionic fields $\psi_i(\boldsymbol k, \tau)$ of the leads while retaining
the
$D(\tau)$ field, we
derive an effective action of the
dot:
\begin{align}
\label{eq:Z}
S_E =&  \int_0^{\hbar\beta} d\tau \Bigg[ \int_0^{\hbar\beta} d\tau' \bar D(\tau) [-G_{dd}^{-1}(\tau,\tau')] D(\tau')
\nonumber\\
& - \frac{C_\Sigma}{2} \left( \frac{\hbar}{2e}i \partial_\tau \phi(\tau)  + \frac{C_I}{2 C_\Sigma} V^a \right)^2 \Bigg].
\end{align}
In the Green's function $-G_{dd}^{-1}(\tau,\tau') = (\hbar\partial_\tau + \epsilon_g  \tau_z)  \delta(\tau-\tau')  + \Sigma(\tau,\tau')$, the first term captures the isolated dot, and the self-energy term
 $\Sigma(\tau,\tau') = \sum_i t_i^2   e^{i \tau_z \frac{\phi_i(\tau)}{2} } \tau_z  g_i(\tau - \tau')  \tau_z e^{-i \tau_z \frac{\phi_i(\tau')}{2} }$ captures the coupling to the leads, where $g_i(\tau) \approx \sum_{n} \left(  - \pi \nu_i\frac{\hbar\omega + \Delta \tau_x }{\sqrt{\Delta^2 - (\hbar\omega)^2}} \right) \frac{ e^{-\omega \tau} }{\hbar\beta}\Big|_{\omega=i \omega_n}$ is the momentum-integrated Green's function for isolated leads per volume
 in the time-domain within a wide-band approximation, $\nu_i$ is the density of states per spin at Fermi level,
 and $\omega_n$ are fermionic Matsubara frequencies.
 We use this non-perturbative result only for the in-gap contributions;
its evaluation in general is
an open problem.

When the ABS bands, $\pm E_A(\phi)$, are well gapped from the continuum
\cite{kringhoj_suppressed_2020-1,bargerbos_observation_2020-1},
and the charging energy is small ($E_C = e^2/2C_\Sigma \ll \Delta$ leading to slow phase dynamics),
the denominators of $g_i(\tau)$ can be approximated adiabatically \cite{zazunov_dynamics_2005}
as $\sqrt{\Delta^2 - (\hbar\omega)^2}\approx \zeta = \zeta(\phi) = \sqrt{\Delta^2 - E_A(\phi)^2}$.
$\zeta$ is treated as a constant within the Matsubara summation provided that the ABS bands are
sufficiently flat, $E_A(\phi) \partial_\phi E_A(\phi) \ll \Delta^2 - E_A^2(\phi)$, which is
the case
in the weak tunneling regime $\Gamma_i \ll \Delta$.
Here, $\pm E_A(\phi)$ are the in-gap ABS levels \cite{beenakker_superconducting_1992} obtained from the poles
of $G_{dd}(\omega)$ in the static limit,  $\partial_{\tau}\phi_i(\tau) \to 0$.
 Within this approximation and at low temperatures,
 $\hbar\beta \gg 1/|\omega|$, $G_{dd}^{-1}(\tau,\tau') \approx G_{dd,\text{a}}^{-1}(\tau) \delta(\tau-\tau')$
 becomes local in time:
\begin{align}
G_{dd,\text{a}}^{-1}(\tau) =  & -\frac{1}{Z_d} \Bigg(   \hbar \partial_\tau
+  Z_d \Bigg[ \sum_{i=L,R} -\frac{\Gamma_i}{ \zeta }    \frac{i\hbar \partial_\tau \phi_i(\tau)}{2} \tau_z
\nonumber\\
& + \frac{\Gamma_i \Delta}{\zeta} e^{i \tau_z \frac{\phi_i(\tau)}{2} }
\tau_x e^{-i \tau_z \frac{\phi_i(\tau)}{2} }
+    \epsilon_g \tau_z \Bigg]
    \Bigg),
\label{eq:Gdda}
\end{align}
where
$\Gamma_i \equiv \pi \nu_i t_i^2$ and $\frac{1}{Z_d} \equiv 1 +  \frac{\Gamma}{\zeta}$ \cite{zeta-footnote}.
Upon substituting this result into Eq.~\eqref{eq:Z}, the first term in the action $S_E$
produces a Hamiltonian that is in agreement
with  Ref. \cite{kurilovich_microwave_2021},
in which
phases were treated as classical parameters and
the $\propto \dot\phi(t)$ term was obtained as a diabatic correction.

\xsection{Contributions of the filled continuum}
We calculate the contribution from the negative continuum energies at zero temperature
perturbatively in $t_i$, in the regime $\Gamma_i \ll \sqrt{\Delta^2 - \epsilon_g^2}$.
At energies $|\hbar\omega| \geq \Delta$, the $D(\tau)$ field is a fast variable and can be integrated out,
along with the fields of the leads $\psi_i(\boldsymbol k, \tau)$
\cite{ambegaokar_quantum_1982,eckern_quantum_1984},
to obtain the leading order contribution from the continuum %
\begin{align}
S_T^{(2)} =
\frac{1}{2} \text{Tr}
\int_0^{\hbar \beta}\int_0^{\hbar \beta} d\tau d\tau'
 & G_0(\tau-\tau') \delta G^{-1}(\tau')
\nonumber\\
 \times &
G_0(\tau'-\tau) \delta G^{-1}(\tau)
\label{eq:continuum-contrib-to-free-energy-2nd-order}
\end{align}
(in time domain). %
Here, $\delta G^{-1}(\tau)$ is the off-diagonal tunneling part of $G^{-1}(\tau)$
in
Eq.~\eqref{eq:Ginv},
and
$G_0(\tau) = \text{diag}(\mathcal V_L g_L(\tau), \mathcal V_R g_R(\tau), g_d(\tau) )$
contains
the
momentum-integrated Green's functions of the uncoupled leads and %
dot.

In order to evaluate $G_0(\tau)$, we first evaluate $g_i(\tau)$.
At low temperatures, $\hbar\beta \gg  \hbar/\Delta$, $g_i(\tau) \approx - \nu_i \Delta \frac{1}{\hbar} \left[\text{sgn}(\tau)K_1\left(|\tau|\frac{\Delta}{\hbar}\right)  + K_0\left(|\tau|\frac{\Delta}{\hbar}\right) \tau_x\right]$ where $K_{0,1}(x)$ are the modified Bessel functions of the second kind. Similarly, $g_d(\tau) = -\frac{1}{\hbar}\text{sgn}(\tau) e^{-\frac{\epsilon_g \tau_z }{\hbar} \tau}$.  Because $K_{0,1}\left(|\tau|\frac{\Delta}{\hbar}\right)$ decay exponentially for $|\tau| \gg \hbar/\Delta$, we
expand the rest of the integrand in the expression for $S_T^{(2)}$ in powers of $\delta\tau = \tau-\tau'$ around $\bar\tau = \frac{\tau+\tau'}{2}$ \cite{eckern_quantum_1984,eckern_effective_2005}. After expanding the phases $\phi_i(\tau)-\phi_i(\tau') = \partial_{\bar\tau}\phi(\bar\tau) \delta\tau + \mathcal O(\delta \tau^3)$ and the exponent containing phases up to second order in $\delta\tau$, we integrate out $\delta\tau$.
The negative continuum contributions to each ABS are extracted as $S_\text{cont}^{(2)} = \frac{1}{2} (S_T^{(2)}[\phi_L,\phi_R,\epsilon_g] - S_T^{(2)}[0,0,0])$:
\begin{widetext}
\begin{align}
\label{eq:S2}
S_\text{cont}^{(2)} &\approx \sum_i  \int_0^{\hbar \beta} d\bar\tau  \left(U^c_i - q^c_i \frac{\hbar}{e} i \partial_{\bar\tau} \phi_i(\bar\tau) +  \frac{C^c_i }{2} \left[ \frac{\hbar}{2e} \partial_{\bar\tau} \phi_i(\bar\tau)\right]^2 \right), \\
U^c_i &= -\Gamma_i \frac{2}{\pi}\epsilon_g \frac{\arcsin\frac{\epsilon_g}{\Delta}}{ \sqrt{\Delta^2 -\epsilon_g^2}}, \qquad
q^c_i = -\Gamma_i e  \frac{\epsilon_g +  \Delta^2 \frac{ \arcsin\frac{\epsilon_g}{\Delta}}{ \sqrt{\Delta^2 -\epsilon_g^2}} }{\pi (\Delta^2 -\epsilon_g^2)}, \qquad
C^c_i = \Gamma_i 2e^2 \frac{2 \Delta^2 + \epsilon_g^2 + 3\Delta^2 \epsilon_g   \frac{\arcsin\frac{\epsilon_g}{\Delta}}{ \sqrt{\Delta^2 -\epsilon_g^2}}   }{ \pi(\Delta^2 -\epsilon_g^2)^2} \nonumber
\end{align}
\end{widetext}
within the slow phase approximation $|i\hbar\partial_{\bar\tau} \phi_i(\bar\tau)| \ll \Delta-|\epsilon_g|$
for $|\epsilon_g| < \Delta$,
and $U^c_i$, $q^c_i$, $C^c_i$ respectively determine the energy shift, charge offset,
and capacitance renormalizations for each lead.
The dynamic contributions, $\propto q^c_i, C^c_i$,
are associated with effects of quantum phase fluctuations, see Eq.~(\ref{eq:Heff}) below.
In a circuit representation, $C^c_i$ is a capacitance that is in parallel to capacitances
between the dot and the leads (like $C_{bi}$ in Fig.~\ref{fig:circuit}).

Higher order corrections in the slow phase approximation become significant as $|\epsilon_g|$
approaches $\Delta$; for typical gatemon values $E_C/h \lesssim 0.5$GHz
 and $\Delta/h \sim 40\text{--}50$GHz,
 \cite{kringhoj_suppressed_2020-1,bargerbos_observation_2020-1},
 the above expression remains adequate for $|\epsilon_g| \lesssim 0.7 \Delta$
 (with a truncation error up to $\approx 5\%$).

$S_T^{(2)}$ corresponds to a sum
of bubble diagrams which can be interpreted
as the creation (at time $\tau$) and annihilation (at time $\tau'$) of virtual particle-hole pairs by
two tunneling events, with one of the pair located at either of the leads experiencing
a potential $\pm e V_i(\tau)$ and the other at the dot experiencing $\mp \epsilon_g$ for
a duration $|\tau'-\tau| \lesssim \frac{\hbar}{\Delta - |\epsilon_g|}$.
This results in the $V_i(\tau)$-dependent (quadratic, due to expansion in $\delta \tau$)
and $\epsilon_g$-dependent contributions to the ABS energies obtained above.

We have so far obtained the leading order correction to the capacitance and charge offset.
The lack of $\phi_i(\tau)$-dependence in $S_\text{cont}^{(2)}$ is expected, since it is of second order in tunneling.
In order to capture the leading order corrections to the supercurrent, we calculate the next order term in the
tunnelings by neglecting phase fluctuations:
$\frac{S_T^{(4)}}{\hbar} =  \sum_n \frac{1}{4}\text{Tr}\left([G_0(i \omega_n) \delta G^{-1}]^4\right)$.
This static treatment disregards next order contributions to the capacitance and charge offset which
come with an additional smallness factor $\sim \Gamma_i/\sqrt{\Delta^2 - \epsilon_g^2}$.
The continuum contribution is $S_\text{cont}^\text{(4)} = \int_0^{\hbar\beta} d\tau \mathcal U^c(\phi(\tau))$ where
\begin{align}
\mathcal U^c(\phi) &=  \frac{ -2\Gamma_L \Gamma_R \Delta^2 \sin^2\frac{\phi}{2}  + \Gamma^2 \epsilon_g^2 \left( 1+\frac{\Delta^2}{\Delta^2-\epsilon_g^2} \right)  }{ \Delta (\Delta^2 - \epsilon_g^2) }.
\end{align}
For larger $\epsilon_g \lesssim \Delta$, the contribution of this term to the supercurrent can become
as important as the in-gap contributions ($\partial_\phi E_A(\phi) \sim \partial_\phi \mathcal U^c(\phi)$).

The combined energy shift defined as the static portions of $S_\text{cont}^\text{(2)}$
and  $S_\text{cont}^\text{(4)}$ is given by $E_\text{cont}(\phi) \equiv \sum_{i=L,R} U^c_i + {\cal U}^c(\phi)$,
which is in numerical agreement with the non-perturbative result given in
Eq.~(12) of Ref.~\cite{kurilovich_microwave_2021} for $\Gamma_i \ll \sqrt{\Delta^2 - \epsilon_g^2}$.
For  $\Gamma_i, \epsilon_g \ll \Delta$,  it simplifies to
\begin{align}
E_\text{cont}(\phi) \approx -\frac{2}{\pi} \Gamma \frac{\epsilon_g^2}{\Delta^2}
-\frac{2 \Gamma_L \Gamma_R}{\Delta} \left(1 + \frac{\epsilon_g^2}{\Delta^2} \right) \sin^2\frac{\phi}{2}
+ \frac{2\Gamma^2 \epsilon_g^2}{\Delta^3}.
\end{align}

 \begin{figure}
    \centering
   \includegraphics[width=1\columnwidth]{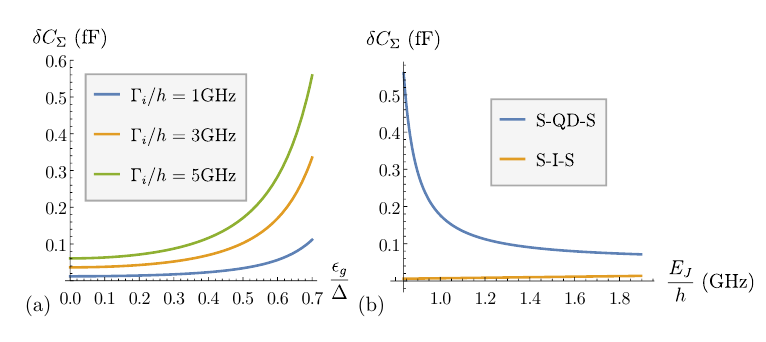}
\caption{(a) Change in effective shunting capacitance across the junction, $\delta C_\Sigma$,
as a function of gate voltage, $\epsilon_g$, for a set of representative values of
$\Gamma_i$ at $\Delta/h = 45 \text{GHz}$.
(b) A comparison of $\delta C_\Sigma$ for S-QD-S junction, Eq.~\eqref{eq:deltaC},
at low transparencies $T$, and the analogous value of $\delta C_{\Sigma}^{\rm JJ} = \frac{3 e^2}{8\Delta^2}E_J$
for an S-I-S junction \cite{eckern_quantum_1984} at given values $E_J$.
For the S-QD-S curve, an effective $E_J$ is defined via Eq.~\eqref{eq:E_J-eff}
as $\epsilon_g/\Delta $ is varied between $0.2$ and $0.7$ for $\Gamma_i/h = 5 \text{GHz}$.
In a typical S-I-S junction used for transmon with many channels, the value of $E_J/h$ is
$\sim 10 \text{GHz}$ \cite{berke2022transmon}.
    }
    \label{fig:capacitance}
\end{figure}

\xsection{Phase quantization: converting path integral to quantum Hamiltonian}
Combining the in-gap and continuum contributions results in
the effective action:
$S_\text{ABS}
= \int_0^{\hbar\beta} d\bar\tau \bar D(\bar\tau) [-G_{dd,\text{a}}^{-1}(\bar\tau)] D(\bar\tau)
+ S_\text{cont}^{(2)} + S_\text{cont}^{(4)} - \frac{C_\Sigma}{2} \left( \frac{\hbar}{2e} i \partial_\tau \phi(\tau)
+ \frac{C_I}{2 C_\Sigma} V^a \right)^2$.
At this point, it is convenient to express the action in terms of difference and average phases $\phi(\bar\tau)$ and
$\phi_{\rm av}(\bar\tau) = \frac{\phi_L(\bar\tau) + \phi_R(\bar\tau)}{2}$.
The weakly coupled $\phi(\tau)$ and
$\phi_{\rm av}(\tau)$
fields can be decoupled perturbatively, and the confinement potential of
$\phi_{\rm av}(\tau)$
can be gauged away.
It can be shown that the partition function obtained from the Hamiltonian
\begin{align}
\label{eq:Heff}
    \hat H &= 4 \tilde E_C\left( \hat n -  \tilde n_g - n_z  \hat D^\dagger \tau_z \hat D   \right)^2  + E_\text{cont}(\phi)  \\
    & + Z_d  \hat D^\dagger \left(     \frac{\Delta}{\zeta} \left[\Gamma \cos\frac{\hat \phi}{2} \tau_x - \delta\Gamma \sin\frac{\hat \phi}{2} \tau_y\right] + \epsilon_g \tau_z   \right)  \hat D, \nonumber
\end{align}
is $Z_G \propto \int \mathcal D\phi e^{-{S_{\rm ABS}}/{\hbar}}$.
The average phase
$\hat{\phi}_{\rm av}(\tau)$
does not appear in $\hat H$, since its conjugate charge operator commutes with
$\hat H$
which allows
$\hat{\phi}_{\rm av}(\tau)$
to be replaced with a constant.
Above, $\hat n$ and $\hat \phi$ are conjugate quantum operators satisfying $[\hat \phi, \hat n] = i$
as a result of the mapping of the functional integration over $\phi(\tau)$ onto operator formalism, and
$\hat D = (\hat d_\uparrow, \hat d^\dagger_\downarrow)^T$ is the dot field operator in Nambu space.
The charging Hamiltonian contains the charging energy
$\tilde E_C = \frac{e^2}{2 (C_\Sigma + \delta C_\Sigma)}$
with
\begin{subequations}
\begin{align}
\label{eq:deltaC}
\delta C_\Sigma(\epsilon_g) = \left[(C_L^c)^{-1} + (C_R^c)^{-1}\right]^{-1},
\\
\label{eq:n_z}
n_z(\epsilon_g) = \frac{C_\Sigma + \frac{C_L^c + C_R^c}{4} }{ C_\Sigma + \delta C_\Sigma }
\frac{\delta\Gamma}{4(\zeta + \Gamma)},\\
\label{eq:dn_g}
\tilde n_g(\epsilon_g) =
\frac{C_\Sigma + \frac{C_L^c + C_R^c}{4} }{ C_\Sigma + \delta C_\Sigma } \left(n_g + \frac{q^c_L - q^c_R}{2 e} \right),
\end{align}
\end{subequations}
where $n_g = \frac{1}{2e} \frac{C_I}{2}V^a$ is the usual charge offset due to
an
applied voltage
(see Fig.~\ref{fig:circuit}),
$n_z(\epsilon_g)$
is the strength of the charge offset term which depends on the dot occupation,
and
$\delta n_g(\epsilon_g)
\simeq \tilde n_g(\epsilon_g) - n_g
\approx \frac{q^c_L - q^c_R}{2 e} $
is the charge offset induced by
the continuum contributions from Eq.~\eqref{eq:S2}, and $ \delta \Gamma = \Gamma_L - \Gamma_R$.
In Fig.~\ref{fig:capacitance}(a), the change in capacitance $\delta C_\Sigma$ is shown as a function
of $\epsilon_g \in [0, 0.7\Delta]$ for a few representative values $\Gamma_i \ll \sqrt{\Delta^2 - \epsilon_g^2}$.
As an example of
the charge offsets
\cite{non-perturbative_nz}
$n_z(\epsilon_g)$ and $\delta n_g(\epsilon_g)$
given by Eqns.~\eqref{eq:n_z} and \eqref{eq:dn_g},
which arise in asymmetric junctions, their values for
$\epsilon_g=0.7 \Delta$, $\Gamma_L/h=5$GHz and $\Gamma_R/h=1$GHz
are respectively $0.02$ and $-0.05$.

Since the parity operator, $(\hat D^\dagger \tau_z \hat D)^2$,
commutes with $\hat H$, the even- and odd-parity sectors are decoupled
\footnote{
By projecting onto the odd parity subspace, $\{|\uparrow\rangle$, $|\downarrow\rangle\}$,
we obtain
$\hat H_\text{odd}=4 \tilde E_C\left(-i \partial_\phi - \tilde n_g  \right)^2 + E_\text{cont}(\phi)$,
which is relevant when the Coulomb interaction is negligible.}.
Projecting Eq.~\eqref{eq:Heff} onto the even occupancy states
$|0\rangle$ and $|\uparrow\downarrow\rangle = d_{\uparrow}^\dagger d_{\downarrow}^\dagger |0\rangle$,
we obtain
\begin{equation}
{\hat H}_\text{even} =  4 {\tilde E}_C \left(-i \partial_\phi - {\tilde n}_g - n_z \eta_z  \right)^2 + \hat U_J(\phi)
\label{eq:Heven}
\end{equation}
where $\hat U_J(\phi)$ is the $2\times2$ Josephson matrix potential,
which includes in-gap (ABS) and continuum contributions,
\begin{align}
\hat U_J(\phi) = &
Z_d \left(\frac{\Delta}{\zeta}\left[\Gamma \cos\frac{\phi}{2} \eta_x - \delta\Gamma \sin\frac{\phi}{2} \eta_y\right]
+ \epsilon_g \eta_z\right)
\nonumber\\
& + E_{\rm cont}(\phi),
\label{eq:U_J}
\end{align}
where $Z_d \equiv \frac{\zeta}{\zeta + \Gamma}$,
and $\eta_{x,y,z}$ are the Pauli matrices acting on the even occupancy space of the dot.
To the second order in $t_i$, qualitatively, for a doubly occupied (unoccupied) dot either
(i) a single electron (hole) can tunnel to a lead and back or (ii) a pair of electrons (holes) can cotunnel
to one of the leads.
Since electrons and holes differ in charge, the first process will generally lead to occupancy dependent
charge offset, since $\propto n_z \eta_z$. The second process corresponds to change of the dot occupancy,
reflected in the off-diagonal terms of Eq.~\eqref{eq:Heven}.
When one of the leads is
disconnected from the dot,
the phase dependence of $\hat H_\text{even}$ can be removed and
the supercurrent vanishes.

To compare to a typical S-I-S JJ we consider a low-transparency regime in Eq.~\eqref{eq:Heven}
where the charging energy will be neglected. The eigenvalues of the potential term $\propto Z_d$ in
$\hat H_\text{even}$ yields the well-known result for the in-gap
ABS energies \cite{beenakker_superconducting_1992}
\begin{align}
E_A(\phi) = \frac{\Delta}{\zeta + \Gamma} \sqrt{\Gamma^2 + \frac{\epsilon_g^2\zeta^2}{\Delta^2}}
\sqrt{1 - T(\epsilon_g) \sin^2\frac{\phi}{2}} ,
\label{eq:E_A-ABS}
\end{align}
where we defined the transparency as
\begin{equation}
T(\epsilon_g) \equiv \frac{4 \Gamma_L \Gamma_R}{ \Gamma^2 + \epsilon_g^2 \frac{\zeta^2}{\Delta^2} }
\label{eq:Transparency} .
\end{equation}
In the regime of small $\Gamma_i \ll \sqrt{\Delta^2 - \epsilon_g^2}$ and small $\epsilon_g$,
$T(\epsilon_g)$ takes the Breit--Wigner form (cf. Ref.~\cite{kringhoj_suppressed_2020-1}).
In the low-transparency limit, $T\ll 1$, (reached either for $\Gamma_L \gg \Gamma_R$ or for relatively
large gate voltages, $\epsilon_g$),
ABS are well-gapped allowing to obtain an effective Josephson energy
\begin{align}
E_{J,{\rm eff}}^A \approx \frac{\Delta}{\zeta + \Gamma}
\frac{\Gamma_L \Gamma_R}{\sqrt{\Gamma^2 + \frac{\epsilon_g^2\zeta^2}{\Delta^2}}} .
\label{eq:E_J-eff}
\end{align}
In Fig.~\ref{fig:capacitance}(b), we use this definition to make a comparison
between $\delta C_{\Sigma}$ for an S-QD-S junction, and
the analogous $\delta C_\Sigma^\text{JJ}$ for an S-I-S junction at a given value
of $E_J$ \cite{eckern_quantum_1984}, which shows that the capacitance renormalization is
one to two orders of magnitude stronger for the S-QD-S junction
\footnote{A more rigorous definition of $E_{J,{\rm eff}}$ with exact continuum
contributions does not alter this result.}.

\xsection{Discussion and outlook}
In this paper we have developed a self-consistent approach,
reducing the underlying many-body system of a S-QD-S junction to a simple Hamiltonian of
a single variable ---the phase difference across the junction.
When the S-QD-S junction is shunted by capacitance $C_{\Sigma}$, the phase $\hat{\phi}$ is a
quantum operator and the associated charging energy $E_C$ gets renormalized as
$\tilde E_C = \frac{e^2}{2(C_{\Sigma} + \delta C_{\Sigma}(\epsilon_g))}$
while the Hamiltonian becomes a $2\times 2$ matrix acting on the even occupancy states of the dot.
In addition to the capacitance, the charge offset gets new QD gate voltage dependent contributions,
$\delta n_g(\epsilon_g)$, $n_z(\epsilon_g)$ that arise in an asymmetric situation, when
hopping
rates to the left and to the right are different, $\Gamma_L \neq \Gamma_R$.

A direct experimental probe of the capacitance renormalization $\delta C_\Sigma$
(as well as $\delta n_g$ and $n_z$) would be to measure changes in the anharmonicity $\alpha$
of
a gatemon while sweeping the gate voltage $\epsilon_g$.
The predicted strength of $\delta C_\Sigma$ and its sensitive dependence on $\epsilon_g$ suggests
that it could be important to account for when designing high-fidelity quantum gates, e.g.,
in architectures utilizing $\epsilon_g$ modulation to realize entangling gates between qubits.

The new charge offsets $\tilde{n}_g$ and $n_z$
that arise for asymmetric tunnelings $\Gamma_L \neq \Gamma_R$,
and their dependence on $\epsilon_g$ and $\Gamma_i$
will be important in the interpretation of previous
\cite{kringhoj_suppressed_2020-1,bargerbos_observation_2020-1,vakhtel_quantum_2023} and
new experiments, e.g., Refs.~\cite{matute2023quantum,PinoSoutoAguado2024},
in this growing field of quantum circuits based on super-semi junctions.
The $\epsilon_g$ and $\Gamma_i$ dependence of the new offset charges $n_z$ and $\delta n_g$ could also be
the basis for coupling super-semi junction circuits to each other or to a general
circuit
environment in new ways.

From a theoretical standpoint, it is desirable to extend our
method to strong tunneling regime, $\Gamma_i \sim \Delta$, e.g. in a non-perturbative approach,
to obtain higher-order corrections to time-dependent continuum contributions,
like $S_\text{cont}^{(4)}$ (that would lead to phase-dependent corrections to the capacitance).
Also note that non-perturbative calculations for
$n_z$ \cite{non-perturbative_nz} and $E_\text{cont}(\phi)$ \cite{kurilovich_microwave_2021,fatemi_nonlinearity_2024}
show they can grow an order magnitude in the strong tunneling regime.
These results are also relevant
to treat multiterminal devices \cite{matute2023quantum},
and to explicitly include multichannel physics
which is particularly relevant for planar super-semi junctions \cite{casparis2018superconducting,PhysRevResearch.6.023094,liu2025strongly}.

\xsection{Acknowledgements}
We acknowledge helpful discussions with Pavel D. Kurilovich, Max Hays,
Thomas Hazard, and Emily Toomey. This research was funded by the LPS Qubit Collaboratory,
and in part under Air Force Contract No. FA8702-15-D-0001.
Any opinions, findings, conclusions or recommendations expressed in this material are those
of the authors and do not necessarily reflect the views of the US Air Force or the US Government.

\bibliography{zotero,extra-rus}

\begin{thebibliography}{44}%
\makeatletter
\providecommand \@ifxundefined [1]{%
 \@ifx{#1\undefined}
}%
\providecommand \@ifnum [1]{%
 \ifnum #1\expandafter \@firstoftwo
 \else \expandafter \@secondoftwo
 \fi
}%
\providecommand \@ifx [1]{%
 \ifx #1\expandafter \@firstoftwo
 \else \expandafter \@secondoftwo
 \fi
}%
\providecommand \natexlab [1]{#1}%
\providecommand \enquote  [1]{``#1''}%
\providecommand \bibnamefont  [1]{#1}%
\providecommand \bibfnamefont [1]{#1}%
\providecommand \citenamefont [1]{#1}%
\providecommand \href@noop [0]{\@secondoftwo}%
\providecommand \href [0]{\begingroup \@sanitize@url \@href}%
\providecommand \@href[1]{\@@startlink{#1}\@@href}%
\providecommand \@@href[1]{\endgroup#1\@@endlink}%
\providecommand \@sanitize@url [0]{\catcode `\\12\catcode `\$12\catcode
  `\&12\catcode `\#12\catcode `\^12\catcode `\_12\catcode `\%12\relax}%
\providecommand \@@startlink[1]{}%
\providecommand \@@endlink[0]{}%
\providecommand \url  [0]{\begingroup\@sanitize@url \@url }%
\providecommand \@url [1]{\endgroup\@href {#1}{\urlprefix }}%
\providecommand \urlprefix  [0]{URL }%
\providecommand \Eprint [0]{\href }%
\providecommand \doibase [0]{https://doi.org/}%
\providecommand \selectlanguage [0]{\@gobble}%
\providecommand \bibinfo  [0]{\@secondoftwo}%
\providecommand \bibfield  [0]{\@secondoftwo}%
\providecommand \translation [1]{[#1]}%
\providecommand \BibitemOpen [0]{}%
\providecommand \bibitemStop [0]{}%
\providecommand \bibitemNoStop [0]{.\EOS\space}%
\providecommand \EOS [0]{\spacefactor3000\relax}%
\providecommand \BibitemShut  [1]{\csname bibitem#1\endcsname}%
\let\auto@bib@innerbib\@empty
\bibitem [{\citenamefont {Kjaergaard}\ \emph {et~al.}(2020)\citenamefont
  {Kjaergaard}, \citenamefont {Schwartz}, \citenamefont {Braumüller},
  \citenamefont {Krantz}, \citenamefont {Wang}, \citenamefont {Gustavsson},\
  and\ \citenamefont {Oliver}}]{kjaergaard_superconducting_2020}%
  \BibitemOpen
  \bibfield  {author} {\bibinfo {author} {\bibfnamefont {M.}~\bibnamefont
  {Kjaergaard}}, \bibinfo {author} {\bibfnamefont {M.~E.}\ \bibnamefont
  {Schwartz}}, \bibinfo {author} {\bibfnamefont {J.}~\bibnamefont
  {Braumüller}}, \bibinfo {author} {\bibfnamefont {P.}~\bibnamefont {Krantz}},
  \bibinfo {author} {\bibfnamefont {J.~I.-J.}\ \bibnamefont {Wang}}, \bibinfo
  {author} {\bibfnamefont {S.}~\bibnamefont {Gustavsson}},\ and\ \bibinfo
  {author} {\bibfnamefont {W.~D.}\ \bibnamefont {Oliver}},\ }\bibfield  {title}
  {\bibinfo {title} {Superconducting {Qubits}: {Current} {State} of {Play}},\
  }\href {https://doi.org/10.1146/annurev-conmatphys-031119-050605} {\bibfield
  {journal} {\bibinfo  {journal} {Annual Review of Condensed Matter Physics}\
  }\textbf {\bibinfo {volume} {11}},\ \bibinfo {pages} {369} (\bibinfo {year}
  {2020})}\BibitemShut {NoStop}%
\bibitem [{\citenamefont {Bravyi}\ \emph {et~al.}(2022)\citenamefont {Bravyi},
  \citenamefont {Dial}, \citenamefont {Gambetta}, \citenamefont {Gil},\ and\
  \citenamefont {Nazario}}]{bravyi_future_2022}%
  \BibitemOpen
  \bibfield  {author} {\bibinfo {author} {\bibfnamefont {S.}~\bibnamefont
  {Bravyi}}, \bibinfo {author} {\bibfnamefont {O.}~\bibnamefont {Dial}},
  \bibinfo {author} {\bibfnamefont {J.~M.}\ \bibnamefont {Gambetta}}, \bibinfo
  {author} {\bibfnamefont {D.}~\bibnamefont {Gil}},\ and\ \bibinfo {author}
  {\bibfnamefont {Z.}~\bibnamefont {Nazario}},\ }\bibfield  {title} {\bibinfo
  {title} {The future of quantum computing with superconducting qubits},\
  }\href {https://doi.org/10.1063/5.0082975} {\bibfield  {journal} {\bibinfo
  {journal} {Journal of Applied Physics}\ }\textbf {\bibinfo {volume} {132}},\
  \bibinfo {pages} {160902} (\bibinfo {year} {2022})}\BibitemShut {NoStop}%
\bibitem [{\citenamefont {Koch}\ \emph {et~al.}(2007)\citenamefont {Koch},
  \citenamefont {Yu}, \citenamefont {Gambetta}, \citenamefont {Houck},
  \citenamefont {Schuster}, \citenamefont {Majer}, \citenamefont {Blais},
  \citenamefont {Devoret}, \citenamefont {Girvin},\ and\ \citenamefont
  {Schoelkopf}}]{koch_charge-insensitive_2007}%
  \BibitemOpen
  \bibfield  {author} {\bibinfo {author} {\bibfnamefont {J.}~\bibnamefont
  {Koch}}, \bibinfo {author} {\bibfnamefont {T.~M.}\ \bibnamefont {Yu}},
  \bibinfo {author} {\bibfnamefont {J.}~\bibnamefont {Gambetta}}, \bibinfo
  {author} {\bibfnamefont {A.~A.}\ \bibnamefont {Houck}}, \bibinfo {author}
  {\bibfnamefont {D.~I.}\ \bibnamefont {Schuster}}, \bibinfo {author}
  {\bibfnamefont {J.}~\bibnamefont {Majer}}, \bibinfo {author} {\bibfnamefont
  {A.}~\bibnamefont {Blais}}, \bibinfo {author} {\bibfnamefont {M.~H.}\
  \bibnamefont {Devoret}}, \bibinfo {author} {\bibfnamefont {S.~M.}\
  \bibnamefont {Girvin}},\ and\ \bibinfo {author} {\bibfnamefont {R.~J.}\
  \bibnamefont {Schoelkopf}},\ }\bibfield  {title} {\bibinfo {title}
  {Charge-insensitive qubit design derived from the {Cooper} pair box},\ }\href
  {https://doi.org/10.1103/PhysRevA.76.042319} {\bibfield  {journal} {\bibinfo
  {journal} {Phys. Rev. A}\ }\textbf {\bibinfo {volume} {76}},\ \bibinfo
  {pages} {042319} (\bibinfo {year} {2007})}\BibitemShut {NoStop}%
\bibitem [{\citenamefont {Mooij}\ \emph {et~al.}(1999)\citenamefont {Mooij},
  \citenamefont {Orlando}, \citenamefont {Levitov}, \citenamefont {Tian},
  \citenamefont {van~der Wal},\ and\ \citenamefont
  {Lloyd}}]{doi:10.1126/science.285.5430.1036}%
  \BibitemOpen
  \bibfield  {author} {\bibinfo {author} {\bibfnamefont {J.~E.}\ \bibnamefont
  {Mooij}}, \bibinfo {author} {\bibfnamefont {T.~P.}\ \bibnamefont {Orlando}},
  \bibinfo {author} {\bibfnamefont {L.}~\bibnamefont {Levitov}}, \bibinfo
  {author} {\bibfnamefont {L.}~\bibnamefont {Tian}}, \bibinfo {author}
  {\bibfnamefont {C.~H.}\ \bibnamefont {van~der Wal}},\ and\ \bibinfo {author}
  {\bibfnamefont {S.}~\bibnamefont {Lloyd}},\ }\bibfield  {title} {\bibinfo
  {title} {Josephson persistent-current qubit},\ }\href
  {https://doi.org/10.1126/science.285.5430.1036} {\bibfield  {journal}
  {\bibinfo  {journal} {Science}\ }\textbf {\bibinfo {volume} {285}},\ \bibinfo
  {pages} {1036} (\bibinfo {year} {1999})}\BibitemShut {NoStop}%
\bibitem [{\citenamefont {Manucharyan}\ \emph {et~al.}(2009)\citenamefont
  {Manucharyan}, \citenamefont {Koch}, \citenamefont {Glazman},\ and\
  \citenamefont {Devoret}}]{manucharyan_fluxonium_2009}%
  \BibitemOpen
  \bibfield  {author} {\bibinfo {author} {\bibfnamefont {V.~E.}\ \bibnamefont
  {Manucharyan}}, \bibinfo {author} {\bibfnamefont {J.}~\bibnamefont {Koch}},
  \bibinfo {author} {\bibfnamefont {L.~I.}\ \bibnamefont {Glazman}},\ and\
  \bibinfo {author} {\bibfnamefont {M.~H.}\ \bibnamefont {Devoret}},\
  }\bibfield  {title} {\bibinfo {title} {Fluxonium: {Single} {Cooper}-{Pair}
  {Circuit} {Free} of {Charge} {Offsets}},\ }\href
  {https://doi.org/10.1126/science.1175552} {\bibfield  {journal} {\bibinfo
  {journal} {Science}\ }\textbf {\bibinfo {volume} {326}},\ \bibinfo {pages}
  {113} (\bibinfo {year} {2009})}\BibitemShut {NoStop}%
\bibitem [{\citenamefont {Gyenis}\ \emph {et~al.}(2021)\citenamefont {Gyenis},
  \citenamefont {Mundada}, \citenamefont {Di~Paolo}, \citenamefont {Hazard},
  \citenamefont {You}, \citenamefont {Schuster}, \citenamefont {Koch},
  \citenamefont {Blais},\ and\ \citenamefont
  {Houck}}]{gyenis_experimental_2021}%
  \BibitemOpen
  \bibfield  {author} {\bibinfo {author} {\bibfnamefont {A.}~\bibnamefont
  {Gyenis}}, \bibinfo {author} {\bibfnamefont {P.~S.}\ \bibnamefont {Mundada}},
  \bibinfo {author} {\bibfnamefont {A.}~\bibnamefont {Di~Paolo}}, \bibinfo
  {author} {\bibfnamefont {T.~M.}\ \bibnamefont {Hazard}}, \bibinfo {author}
  {\bibfnamefont {X.}~\bibnamefont {You}}, \bibinfo {author} {\bibfnamefont
  {D.~I.}\ \bibnamefont {Schuster}}, \bibinfo {author} {\bibfnamefont
  {J.}~\bibnamefont {Koch}}, \bibinfo {author} {\bibfnamefont {A.}~\bibnamefont
  {Blais}},\ and\ \bibinfo {author} {\bibfnamefont {A.~A.}\ \bibnamefont
  {Houck}},\ }\bibfield  {title} {\bibinfo {title} {Experimental {Realization}
  of a {Protected} {Superconducting} {Circuit} {Derived} from the $0$--$\pi$
  {Qubit}},\ }\href {https://doi.org/10.1103/PRXQuantum.2.010339} {\bibfield
  {journal} {\bibinfo  {journal} {PRX Quantum}\ }\textbf {\bibinfo {volume}
  {2}},\ \bibinfo {pages} {010339} (\bibinfo {year} {2021})}\BibitemShut
  {NoStop}%
\bibitem [{\citenamefont {Larsen}\ \emph {et~al.}(2015)\citenamefont {Larsen},
  \citenamefont {Petersson}, \citenamefont {Kuemmeth}, \citenamefont
  {Jespersen}, \citenamefont {Krogstrup}, \citenamefont {Nygård},\ and\
  \citenamefont {Marcus}}]{larsen_semiconductor-nanowire-based_2015}%
  \BibitemOpen
  \bibfield  {author} {\bibinfo {author} {\bibfnamefont {T.}~\bibnamefont
  {Larsen}}, \bibinfo {author} {\bibfnamefont {K.}~\bibnamefont {Petersson}},
  \bibinfo {author} {\bibfnamefont {F.}~\bibnamefont {Kuemmeth}}, \bibinfo
  {author} {\bibfnamefont {T.}~\bibnamefont {Jespersen}}, \bibinfo {author}
  {\bibfnamefont {P.}~\bibnamefont {Krogstrup}}, \bibinfo {author}
  {\bibfnamefont {J.}~\bibnamefont {Nygård}},\ and\ \bibinfo {author}
  {\bibfnamefont {C.}~\bibnamefont {Marcus}},\ }\bibfield  {title} {\bibinfo
  {title} {Semiconductor-{Nanowire}-{Based} {Superconducting} {Qubit}},\ }\href
  {https://doi.org/10.1103/PhysRevLett.115.127001} {\bibfield  {journal}
  {\bibinfo  {journal} {Phys. Rev. Lett.}\ }\textbf {\bibinfo {volume} {115}},\
  \bibinfo {pages} {127001} (\bibinfo {year} {2015})}\BibitemShut {NoStop}%
\bibitem [{\citenamefont {de~Lange}\ \emph {et~al.}(2015)\citenamefont
  {de~Lange}, \citenamefont {van Heck}, \citenamefont {Bruno}, \citenamefont
  {van Woerkom}, \citenamefont {Geresdi}, \citenamefont {Plissard},
  \citenamefont {Bakkers}, \citenamefont {Akhmerov},\ and\ \citenamefont
  {DiCarlo}}]{de_lange_realization_2015}%
  \BibitemOpen
  \bibfield  {author} {\bibinfo {author} {\bibfnamefont {G.}~\bibnamefont
  {de~Lange}}, \bibinfo {author} {\bibfnamefont {B.}~\bibnamefont {van Heck}},
  \bibinfo {author} {\bibfnamefont {A.}~\bibnamefont {Bruno}}, \bibinfo
  {author} {\bibfnamefont {D.}~\bibnamefont {van Woerkom}}, \bibinfo {author}
  {\bibfnamefont {A.}~\bibnamefont {Geresdi}}, \bibinfo {author} {\bibfnamefont
  {S.}~\bibnamefont {Plissard}}, \bibinfo {author} {\bibfnamefont
  {E.}~\bibnamefont {Bakkers}}, \bibinfo {author} {\bibfnamefont
  {A.}~\bibnamefont {Akhmerov}},\ and\ \bibinfo {author} {\bibfnamefont
  {L.}~\bibnamefont {DiCarlo}},\ }\bibfield  {title} {\bibinfo {title}
  {Realization of {Microwave} {Quantum} {Circuits} {Using} {Hybrid}
  {Superconducting}-{Semiconducting} {Nanowire} {Josephson} {Elements}},\
  }\href {https://doi.org/10.1103/PhysRevLett.115.127002} {\bibfield  {journal}
  {\bibinfo  {journal} {Phys. Rev. Lett.}\ }\textbf {\bibinfo {volume} {115}},\
  \bibinfo {pages} {127002} (\bibinfo {year} {2015})}\BibitemShut {NoStop}%
\bibitem [{\citenamefont {Pita-Vidal}\ \emph {et~al.}(2020)\citenamefont
  {Pita-Vidal}, \citenamefont {Bargerbos}, \citenamefont {Yang}, \citenamefont
  {van Woerkom}, \citenamefont {Pfaff}, \citenamefont {Haider}, \citenamefont
  {Krogstrup}, \citenamefont {Kouwenhoven}, \citenamefont {de~Lange},\ and\
  \citenamefont {Kou}}]{pita-vidal_gate-tunable_2020}%
  \BibitemOpen
  \bibfield  {author} {\bibinfo {author} {\bibfnamefont {M.}~\bibnamefont
  {Pita-Vidal}}, \bibinfo {author} {\bibfnamefont {A.}~\bibnamefont
  {Bargerbos}}, \bibinfo {author} {\bibfnamefont {C.-K.}\ \bibnamefont {Yang}},
  \bibinfo {author} {\bibfnamefont {D.~J.}\ \bibnamefont {van Woerkom}},
  \bibinfo {author} {\bibfnamefont {W.}~\bibnamefont {Pfaff}}, \bibinfo
  {author} {\bibfnamefont {N.}~\bibnamefont {Haider}}, \bibinfo {author}
  {\bibfnamefont {P.}~\bibnamefont {Krogstrup}}, \bibinfo {author}
  {\bibfnamefont {L.~P.}\ \bibnamefont {Kouwenhoven}}, \bibinfo {author}
  {\bibfnamefont {G.}~\bibnamefont {de~Lange}},\ and\ \bibinfo {author}
  {\bibfnamefont {A.}~\bibnamefont {Kou}},\ }\bibfield  {title} {\bibinfo
  {title} {Gate-{Tunable} {Field}-{Compatible} {Fluxonium}},\ }\href
  {https://doi.org/10.1103/PhysRevApplied.14.064038} {\bibfield  {journal}
  {\bibinfo  {journal} {Phys. Rev. Applied}\ }\textbf {\bibinfo {volume}
  {14}},\ \bibinfo {pages} {064038} (\bibinfo {year} {2020})}\BibitemShut
  {NoStop}%
\bibitem [{\citenamefont {Janvier}\ \emph {et~al.}(2015)\citenamefont
  {Janvier}, \citenamefont {Tosi}, \citenamefont {Bretheau}, \citenamefont
  {Girit}, \citenamefont {Stern}, \citenamefont {Bertet}, \citenamefont
  {Joyez}, \citenamefont {Vion}, \citenamefont {Esteve}, \citenamefont
  {Goffman}, \citenamefont {Pothier},\ and\ \citenamefont
  {Urbina}}]{janvier_coherent_2015}%
  \BibitemOpen
  \bibfield  {author} {\bibinfo {author} {\bibfnamefont {C.}~\bibnamefont
  {Janvier}}, \bibinfo {author} {\bibfnamefont {L.}~\bibnamefont {Tosi}},
  \bibinfo {author} {\bibfnamefont {L.}~\bibnamefont {Bretheau}}, \bibinfo
  {author} {\bibfnamefont {C.~O.}\ \bibnamefont {Girit}}, \bibinfo {author}
  {\bibfnamefont {M.}~\bibnamefont {Stern}}, \bibinfo {author} {\bibfnamefont
  {P.}~\bibnamefont {Bertet}}, \bibinfo {author} {\bibfnamefont
  {P.}~\bibnamefont {Joyez}}, \bibinfo {author} {\bibfnamefont
  {D.}~\bibnamefont {Vion}}, \bibinfo {author} {\bibfnamefont {D.}~\bibnamefont
  {Esteve}}, \bibinfo {author} {\bibfnamefont {M.~F.}\ \bibnamefont {Goffman}},
  \bibinfo {author} {\bibfnamefont {H.}~\bibnamefont {Pothier}},\ and\ \bibinfo
  {author} {\bibfnamefont {C.}~\bibnamefont {Urbina}},\ }\bibfield  {title}
  {\bibinfo {title} {Coherent manipulation of {Andreev} states in
  superconducting atomic contacts},\ }\href
  {https://doi.org/10.1126/science.aab2179} {\bibfield  {journal} {\bibinfo
  {journal} {Science}\ }\textbf {\bibinfo {volume} {349}},\ \bibinfo {pages}
  {1199} (\bibinfo {year} {2015})}\BibitemShut {NoStop}%
\bibitem [{\citenamefont {Hays}\ \emph {et~al.}(2018)\citenamefont {Hays},
  \citenamefont {de~Lange}, \citenamefont {Serniak}, \citenamefont {van
  Woerkom}, \citenamefont {Bouman}, \citenamefont {Krogstrup}, \citenamefont
  {Nygård}, \citenamefont {Geresdi},\ and\ \citenamefont
  {Devoret}}]{hays_direct_2018}%
  \BibitemOpen
  \bibfield  {author} {\bibinfo {author} {\bibfnamefont {M.}~\bibnamefont
  {Hays}}, \bibinfo {author} {\bibfnamefont {G.}~\bibnamefont {de~Lange}},
  \bibinfo {author} {\bibfnamefont {K.}~\bibnamefont {Serniak}}, \bibinfo
  {author} {\bibfnamefont {D.}~\bibnamefont {van Woerkom}}, \bibinfo {author}
  {\bibfnamefont {D.}~\bibnamefont {Bouman}}, \bibinfo {author} {\bibfnamefont
  {P.}~\bibnamefont {Krogstrup}}, \bibinfo {author} {\bibfnamefont
  {J.}~\bibnamefont {Nygård}}, \bibinfo {author} {\bibfnamefont
  {A.}~\bibnamefont {Geresdi}},\ and\ \bibinfo {author} {\bibfnamefont
  {M.}~\bibnamefont {Devoret}},\ }\bibfield  {title} {\bibinfo {title} {Direct
  {Microwave} {Measurement} of {Andreev}-{Bound}-{State} {Dynamics} in a
  {Semiconductor}-{Nanowire} {Josephson} {Junction}},\ }\href
  {https://doi.org/10.1103/PhysRevLett.121.047001} {\bibfield  {journal}
  {\bibinfo  {journal} {Phys. Rev. Lett.}\ }\textbf {\bibinfo {volume} {121}},\
  \bibinfo {pages} {047001} (\bibinfo {year} {2018})}\BibitemShut {NoStop}%
\bibitem [{\citenamefont {Hays}\ \emph {et~al.}(2021)\citenamefont {Hays},
  \citenamefont {Fatemi}, \citenamefont {Bouman}, \citenamefont {Cerrillo},
  \citenamefont {Diamond}, \citenamefont {Serniak}, \citenamefont {Connolly},
  \citenamefont {Krogstrup}, \citenamefont {Nygård}, \citenamefont
  {Levy~Yeyati}, \citenamefont {Geresdi},\ and\ \citenamefont
  {Devoret}}]{hays_coherent_2021-1}%
  \BibitemOpen
  \bibfield  {author} {\bibinfo {author} {\bibfnamefont {M.}~\bibnamefont
  {Hays}}, \bibinfo {author} {\bibfnamefont {V.}~\bibnamefont {Fatemi}},
  \bibinfo {author} {\bibfnamefont {D.}~\bibnamefont {Bouman}}, \bibinfo
  {author} {\bibfnamefont {J.}~\bibnamefont {Cerrillo}}, \bibinfo {author}
  {\bibfnamefont {S.}~\bibnamefont {Diamond}}, \bibinfo {author} {\bibfnamefont
  {K.}~\bibnamefont {Serniak}}, \bibinfo {author} {\bibfnamefont
  {T.}~\bibnamefont {Connolly}}, \bibinfo {author} {\bibfnamefont
  {P.}~\bibnamefont {Krogstrup}}, \bibinfo {author} {\bibfnamefont
  {J.}~\bibnamefont {Nygård}}, \bibinfo {author} {\bibfnamefont
  {A.}~\bibnamefont {Levy~Yeyati}}, \bibinfo {author} {\bibfnamefont
  {A.}~\bibnamefont {Geresdi}},\ and\ \bibinfo {author} {\bibfnamefont {M.~H.}\
  \bibnamefont {Devoret}},\ }\bibfield  {title} {\bibinfo {title} {Coherent
  manipulation of an {Andreev} spin qubit},\ }\href
  {https://doi.org/10.1126/science.abf0345} {\bibfield  {journal} {\bibinfo
  {journal} {Science}\ }\textbf {\bibinfo {volume} {373}},\ \bibinfo {pages}
  {430} (\bibinfo {year} {2021})}\BibitemShut {NoStop}%
\bibitem [{\citenamefont {Pita-Vidal}\ \emph {et~al.}(2023)\citenamefont
  {Pita-Vidal}, \citenamefont {Bargerbos}, \citenamefont {Žitko},
  \citenamefont {Splitthoff}, \citenamefont {Grünhaupt}, \citenamefont
  {Wesdorp}, \citenamefont {Liu}, \citenamefont {Kouwenhoven}, \citenamefont
  {Aguado}, \citenamefont {van Heck}, \citenamefont {Kou},\ and\ \citenamefont
  {Andersen}}]{pita-vidal_direct_2023}%
  \BibitemOpen
  \bibfield  {author} {\bibinfo {author} {\bibfnamefont {M.}~\bibnamefont
  {Pita-Vidal}}, \bibinfo {author} {\bibfnamefont {A.}~\bibnamefont
  {Bargerbos}}, \bibinfo {author} {\bibfnamefont {R.}~\bibnamefont {Žitko}},
  \bibinfo {author} {\bibfnamefont {L.~J.}\ \bibnamefont {Splitthoff}},
  \bibinfo {author} {\bibfnamefont {L.}~\bibnamefont {Grünhaupt}}, \bibinfo
  {author} {\bibfnamefont {J.~J.}\ \bibnamefont {Wesdorp}}, \bibinfo {author}
  {\bibfnamefont {Y.}~\bibnamefont {Liu}}, \bibinfo {author} {\bibfnamefont
  {L.~P.}\ \bibnamefont {Kouwenhoven}}, \bibinfo {author} {\bibfnamefont
  {R.}~\bibnamefont {Aguado}}, \bibinfo {author} {\bibfnamefont
  {B.}~\bibnamefont {van Heck}}, \bibinfo {author} {\bibfnamefont
  {A.}~\bibnamefont {Kou}},\ and\ \bibinfo {author} {\bibfnamefont {C.~K.}\
  \bibnamefont {Andersen}},\ }\bibfield  {title} {\bibinfo {title} {Direct
  manipulation of a superconducting spin qubit strongly coupled to a transmon
  qubit},\ }\href {https://doi.org/10.1038/s41567-023-02071-x} {\bibfield
  {journal} {\bibinfo  {journal} {Nat. Phys.}\ }\textbf {\bibinfo {volume}
  {19}},\ \bibinfo {pages} {1110} (\bibinfo {year} {2023})}\BibitemShut
  {NoStop}%
\bibitem [{\citenamefont {Matute-Ca\~nadas}\ \emph {et~al.}(2024)\citenamefont
  {Matute-Ca\~nadas}, \citenamefont {Tosi},\ and\ \citenamefont
  {Yeyati}}]{matute2023quantum}%
  \BibitemOpen
  \bibfield  {author} {\bibinfo {author} {\bibfnamefont {F.}~\bibnamefont
  {Matute-Ca\~nadas}}, \bibinfo {author} {\bibfnamefont {L.}~\bibnamefont
  {Tosi}},\ and\ \bibinfo {author} {\bibfnamefont {A.~L.}\ \bibnamefont
  {Yeyati}},\ }\bibfield  {title} {\bibinfo {title} {Quantum circuits with
  multiterminal {Josephson-Andreev} junctions},\ }\href
  {https://doi.org/10.1103/PRXQuantum.5.020340} {\bibfield  {journal} {\bibinfo
   {journal} {PRX Quantum}\ }\textbf {\bibinfo {volume} {5}},\ \bibinfo {pages}
  {020340} (\bibinfo {year} {2024})}\BibitemShut {NoStop}%
\bibitem [{\citenamefont {Pino}\ \emph {et~al.}(2024)\citenamefont {Pino},
  \citenamefont {Souto},\ and\ \citenamefont {Aguado}}]{PinoSoutoAguado2024}%
  \BibitemOpen
  \bibfield  {author} {\bibinfo {author} {\bibfnamefont {D.~M.}\ \bibnamefont
  {Pino}}, \bibinfo {author} {\bibfnamefont {R.~S.}\ \bibnamefont {Souto}},\
  and\ \bibinfo {author} {\bibfnamefont {R.}~\bibnamefont {Aguado}},\
  }\bibfield  {title} {\bibinfo {title} {{M}inimal {K}itaev–transmon qubit
  based on double quantum dots},\ }\href
  {https://doi.org/https://doi.org/10.1103/PhysRevB.109.075101} {\bibfield
  {journal} {\bibinfo  {journal} {Phys. Rev. B}\ }\textbf {\bibinfo {volume}
  {109}},\ \bibinfo {pages} {075101} (\bibinfo {year} {2024})}\BibitemShut
  {NoStop}%
\bibitem [{\citenamefont {Beenakker}\ and\ \citenamefont {van
  Houten}(1991)}]{beenakker_josephson_1991}%
  \BibitemOpen
  \bibfield  {author} {\bibinfo {author} {\bibfnamefont {C.~W.~J.}\
  \bibnamefont {Beenakker}}\ and\ \bibinfo {author} {\bibfnamefont
  {H.}~\bibnamefont {van Houten}},\ }\bibfield  {title} {\bibinfo {title}
  {Josephson current through a superconducting quantum point contact shorter
  than the coherence length},\ }\href
  {https://doi.org/10.1103/PhysRevLett.66.3056} {\bibfield  {journal} {\bibinfo
   {journal} {Phys. Rev. Lett.}\ }\textbf {\bibinfo {volume} {66}},\ \bibinfo
  {pages} {3056} (\bibinfo {year} {1991})}\BibitemShut {NoStop}%
\bibitem [{\citenamefont {Zazunov}\ \emph {et~al.}(2003)\citenamefont
  {Zazunov}, \citenamefont {Shumeiko}, \citenamefont {Bratus’}, \citenamefont
  {Lantz},\ and\ \citenamefont {Wendin}}]{zazunov_andreev_2003}%
  \BibitemOpen
  \bibfield  {author} {\bibinfo {author} {\bibfnamefont {A.}~\bibnamefont
  {Zazunov}}, \bibinfo {author} {\bibfnamefont {V.~S.}\ \bibnamefont
  {Shumeiko}}, \bibinfo {author} {\bibfnamefont {E.~N.}\ \bibnamefont
  {Bratus’}}, \bibinfo {author} {\bibfnamefont {J.}~\bibnamefont {Lantz}},\
  and\ \bibinfo {author} {\bibfnamefont {G.}~\bibnamefont {Wendin}},\
  }\bibfield  {title} {\bibinfo {title} {Andreev {Level} {Qubit}},\ }\href
  {https://doi.org/10.1103/PhysRevLett.90.087003} {\bibfield  {journal}
  {\bibinfo  {journal} {Phys. Rev. Lett.}\ }\textbf {\bibinfo {volume} {90}},\
  \bibinfo {pages} {087003} (\bibinfo {year} {2003})}\BibitemShut {NoStop}%
\bibitem [{\citenamefont {Kurilovich}\ \emph {et~al.}(2021)\citenamefont
  {Kurilovich}, \citenamefont {Kurilovich}, \citenamefont {Fatemi},
  \citenamefont {Devoret},\ and\ \citenamefont
  {Glazman}}]{kurilovich_microwave_2021}%
  \BibitemOpen
  \bibfield  {author} {\bibinfo {author} {\bibfnamefont {P.~D.}\ \bibnamefont
  {Kurilovich}}, \bibinfo {author} {\bibfnamefont {V.~D.}\ \bibnamefont
  {Kurilovich}}, \bibinfo {author} {\bibfnamefont {V.}~\bibnamefont {Fatemi}},
  \bibinfo {author} {\bibfnamefont {M.~H.}\ \bibnamefont {Devoret}},\ and\
  \bibinfo {author} {\bibfnamefont {L.~I.}\ \bibnamefont {Glazman}},\
  }\bibfield  {title} {\bibinfo {title} {Microwave response of an {Andreev}
  bound state},\ }\href {https://doi.org/10.1103/PhysRevB.104.174517}
  {\bibfield  {journal} {\bibinfo  {journal} {Phys. Rev. B}\ }\textbf {\bibinfo
  {volume} {104}},\ \bibinfo {pages} {174517} (\bibinfo {year}
  {2021})}\BibitemShut {NoStop}%
\bibitem [{\citenamefont {Likharev}(1986)}]{Likharev-book-1986}%
  \BibitemOpen
  \bibfield  {author} {\bibinfo {author} {\bibfnamefont {K.~K.}\ \bibnamefont
  {Likharev}},\ }\href@noop {} {\emph {\bibinfo {title} {{D}ynamics of
  {J}osephson {J}unctions and {C}ircuits}}}\ (\bibinfo  {publisher} {CRC press,
  Taylor \& Francis group, 1986},\ \bibinfo {year} {1986})\BibitemShut
  {NoStop}%
\bibitem [{\citenamefont {Devoret}(1997)}]{Devoret1997-Les_Houches}%
  \BibitemOpen
  \bibfield  {author} {\bibinfo {author} {\bibfnamefont {M.~H.}\ \bibnamefont
  {Devoret}},\ }\bibinfo {title} {{Q}uantum fluctuations in electrical
  circuits},\ in\ \href@noop {} {\emph {\bibinfo {booktitle} {Quantum
  Fluctuations: Les Houches Session LXIII, June 27 to July 28 1995}}},\
  \bibinfo {editor} {edited by\ \bibinfo {editor} {\bibfnamefont {E.~G.}\
  \bibnamefont {S.~Reynaud}}\ and\ \bibinfo {editor} {\bibfnamefont
  {J.}~\bibnamefont {Zinn-Justin}}}\ (\bibinfo  {publisher} {Elsevier},\
  \bibinfo {address} {Amsterdam},\ \bibinfo {year} {1997})\ pp.\ \bibinfo
  {pages} {351--386}\BibitemShut {NoStop}%
\bibitem [{\citenamefont {Ambegaokar}\ \emph {et~al.}(1982)\citenamefont
  {Ambegaokar}, \citenamefont {Eckern},\ and\ \citenamefont
  {Sch\"{o}n}}]{ambegaokar_quantum_1982}%
  \BibitemOpen
  \bibfield  {author} {\bibinfo {author} {\bibfnamefont {V.}~\bibnamefont
  {Ambegaokar}}, \bibinfo {author} {\bibfnamefont {U.}~\bibnamefont {Eckern}},\
  and\ \bibinfo {author} {\bibfnamefont {G.}~\bibnamefont {Sch\"{o}n}},\
  }\bibfield  {title} {\bibinfo {title} {Quantum {Dynamics} of {Tunneling}
  between {Superconductors}},\ }\href
  {https://doi.org/10.1103/PhysRevLett.48.1745} {\bibfield  {journal} {\bibinfo
   {journal} {Phys. Rev. Lett.}\ }\textbf {\bibinfo {volume} {48}},\ \bibinfo
  {pages} {1745} (\bibinfo {year} {1982})}\BibitemShut {NoStop}%
\bibitem [{\citenamefont {Eckern}\ \emph {et~al.}(1984)\citenamefont {Eckern},
  \citenamefont {Sch\"{o}n},\ and\ \citenamefont
  {Ambegaokar}}]{eckern_quantum_1984}%
  \BibitemOpen
  \bibfield  {author} {\bibinfo {author} {\bibfnamefont {U.}~\bibnamefont
  {Eckern}}, \bibinfo {author} {\bibfnamefont {G.}~\bibnamefont {Sch\"{o}n}},\
  and\ \bibinfo {author} {\bibfnamefont {V.}~\bibnamefont {Ambegaokar}},\
  }\bibfield  {title} {\bibinfo {title} {Quantum dynamics of a superconducting
  tunnel junction},\ }\href {https://doi.org/10.1103/PhysRevB.30.6419}
  {\bibfield  {journal} {\bibinfo  {journal} {Phys. Rev. B}\ }\textbf {\bibinfo
  {volume} {30}},\ \bibinfo {pages} {6419} (\bibinfo {year}
  {1984})}\BibitemShut {NoStop}%
\bibitem [{\citenamefont {Larkin}\ and\ \citenamefont
  {Ovchinnikov}(1983)}]{larkin_decay_1983}%
  \BibitemOpen
  \bibfield  {author} {\bibinfo {author} {\bibfnamefont {A.~I.}\ \bibnamefont
  {Larkin}}\ and\ \bibinfo {author} {\bibfnamefont {Y.~N.}\ \bibnamefont
  {Ovchinnikov}},\ }\bibfield  {title} {\bibinfo {title} {Decay of the
  supercurrent in tunnel junctions},\ }\href
  {https://doi.org/10.1103/PhysRevB.28.6281} {\bibfield  {journal} {\bibinfo
  {journal} {Phys. Rev. B}\ }\textbf {\bibinfo {volume} {28}},\ \bibinfo
  {pages} {6281} (\bibinfo {year} {1983})}\BibitemShut {NoStop}%
\bibitem [{\citenamefont {Kringhøj}\ \emph {et~al.}(2020)\citenamefont
  {Kringhøj}, \citenamefont {van Heck}, \citenamefont {Larsen}, \citenamefont
  {Erlandsson}, \citenamefont {Sabonis}, \citenamefont {Krogstrup},
  \citenamefont {Casparis}, \citenamefont {Petersson},\ and\ \citenamefont
  {Marcus}}]{kringhoj_suppressed_2020-1}%
  \BibitemOpen
  \bibfield  {author} {\bibinfo {author} {\bibfnamefont {A.}~\bibnamefont
  {Kringhøj}}, \bibinfo {author} {\bibfnamefont {B.}~\bibnamefont {van Heck}},
  \bibinfo {author} {\bibfnamefont {T.}~\bibnamefont {Larsen}}, \bibinfo
  {author} {\bibfnamefont {O.}~\bibnamefont {Erlandsson}}, \bibinfo {author}
  {\bibfnamefont {D.}~\bibnamefont {Sabonis}}, \bibinfo {author} {\bibfnamefont
  {P.}~\bibnamefont {Krogstrup}}, \bibinfo {author} {\bibfnamefont
  {L.}~\bibnamefont {Casparis}}, \bibinfo {author} {\bibfnamefont
  {K.}~\bibnamefont {Petersson}},\ and\ \bibinfo {author} {\bibfnamefont
  {C.}~\bibnamefont {Marcus}},\ }\bibfield  {title} {\bibinfo {title}
  {Suppressed {Charge} {Dispersion} via {Resonant} {Tunneling} in a
  {Single}-{Channel} {Transmon}},\ }\href
  {https://doi.org/10.1103/PhysRevLett.124.246803} {\bibfield  {journal}
  {\bibinfo  {journal} {Phys. Rev. Lett.}\ }\textbf {\bibinfo {volume} {124}},\
  \bibinfo {pages} {246803} (\bibinfo {year} {2020})}\BibitemShut {NoStop}%
\bibitem [{\citenamefont {Bargerbos}\ \emph {et~al.}(2020)\citenamefont
  {Bargerbos}, \citenamefont {Uilhoorn}, \citenamefont {Yang}, \citenamefont
  {Krogstrup}, \citenamefont {Kouwenhoven}, \citenamefont {de~Lange},
  \citenamefont {van Heck},\ and\ \citenamefont
  {Kou}}]{bargerbos_observation_2020-1}%
  \BibitemOpen
  \bibfield  {author} {\bibinfo {author} {\bibfnamefont {A.}~\bibnamefont
  {Bargerbos}}, \bibinfo {author} {\bibfnamefont {W.}~\bibnamefont {Uilhoorn}},
  \bibinfo {author} {\bibfnamefont {C.-K.}\ \bibnamefont {Yang}}, \bibinfo
  {author} {\bibfnamefont {P.}~\bibnamefont {Krogstrup}}, \bibinfo {author}
  {\bibfnamefont {L.~P.}\ \bibnamefont {Kouwenhoven}}, \bibinfo {author}
  {\bibfnamefont {G.}~\bibnamefont {de~Lange}}, \bibinfo {author}
  {\bibfnamefont {B.}~\bibnamefont {van Heck}},\ and\ \bibinfo {author}
  {\bibfnamefont {A.}~\bibnamefont {Kou}},\ }\bibfield  {title} {\bibinfo
  {title} {Observation of {Vanishing} {Charge} {Dispersion} of a {Nearly}
  {Open} {Superconducting} {Island}},\ }\href
  {https://doi.org/10.1103/PhysRevLett.124.246802} {\bibfield  {journal}
  {\bibinfo  {journal} {Phys. Rev. Lett.}\ }\textbf {\bibinfo {volume} {124}},\
  \bibinfo {pages} {246802} (\bibinfo {year} {2020})}\BibitemShut {NoStop}%
\bibitem [{\citenamefont {Vakhtel}\ and\ \citenamefont
  {Van~Heck}(2023)}]{vakhtel_quantum_2023}%
  \BibitemOpen
  \bibfield  {author} {\bibinfo {author} {\bibfnamefont {T.}~\bibnamefont
  {Vakhtel}}\ and\ \bibinfo {author} {\bibfnamefont {B.}~\bibnamefont
  {Van~Heck}},\ }\bibfield  {title} {\bibinfo {title} {Quantum phase slips in a
  resonant {Josephson} junction},\ }\href
  {https://doi.org/10.1103/PhysRevB.107.195405} {\bibfield  {journal} {\bibinfo
   {journal} {Phys. Rev. B}\ }\textbf {\bibinfo {volume} {107}},\ \bibinfo
  {pages} {195405} (\bibinfo {year} {2023})}\BibitemShut {NoStop}%
\bibitem [{Note1()}]{Note1}%
  \BibitemOpen
  \bibinfo {note} {Assuming the dot's energy quantization provides the largest
  energy scale, $\delta E \gg \Delta , U$, scf. \cite
  {kurilovich_microwave_2021}}\BibitemShut {NoStop}%
\bibitem [{\citenamefont {Coleman}(2015)}]{coleman_introduction_2015}%
  \BibitemOpen
  \bibfield  {author} {\bibinfo {author} {\bibfnamefont {P.}~\bibnamefont
  {Coleman}},\ }\href {https://doi.org/10.1017/CBO9781139020916} {\emph
  {\bibinfo {title} {Introduction to {Many}-{Body} {Physics}}}}\ (\bibinfo
  {publisher} {Cambridge University Press},\ \bibinfo {address} {Cambridge},\
  \bibinfo {year} {2015})\BibitemShut {NoStop}%
\bibitem [{\citenamefont {Altland}\ and\ \citenamefont
  {Simons}(2010)}]{altland_condensed_2010}%
  \BibitemOpen
  \bibfield  {author} {\bibinfo {author} {\bibfnamefont {A.}~\bibnamefont
  {Altland}}\ and\ \bibinfo {author} {\bibfnamefont {B.}~\bibnamefont
  {Simons}},\ }\href {https://doi.org/10.1017/cbo9780511789984} {\emph
  {\bibinfo {title} {Condensed {Matter} {Field} {Theory}, {Second}
  {Edition}}}},\ \bibinfo {edition} {2nd}\ ed.\ (\bibinfo  {publisher}
  {Cambridge University Press},\ \bibinfo {address} {Cambridge},\ \bibinfo
  {year} {2010})\BibitemShut {NoStop}%
\bibitem [{\citenamefont {Rozhkov}\ and\ \citenamefont
  {Arovas}(1999)}]{rozhkov_josephson_1999}%
  \BibitemOpen
  \bibfield  {author} {\bibinfo {author} {\bibfnamefont {A.~V.}\ \bibnamefont
  {Rozhkov}}\ and\ \bibinfo {author} {\bibfnamefont {D.~P.}\ \bibnamefont
  {Arovas}},\ }\bibfield  {title} {\bibinfo {title} {Josephson {Coupling}
  through a {Magnetic} {Impurity}},\ }\href
  {https://doi.org/10.1103/PhysRevLett.82.2788} {\bibfield  {journal} {\bibinfo
   {journal} {Phys. Rev. Lett.}\ }\textbf {\bibinfo {volume} {82}},\ \bibinfo
  {pages} {2788} (\bibinfo {year} {1999})}\BibitemShut {NoStop}%
\bibitem [{\citenamefont {Meden}(2019)}]{meden_andersonjosephson_2019}%
  \BibitemOpen
  \bibfield  {author} {\bibinfo {author} {\bibfnamefont {V.}~\bibnamefont
  {Meden}},\ }\bibfield  {title} {\bibinfo {title} {The
  {Anderson}–{Josephson} quantum dot—a theory perspective},\ }\href
  {https://doi.org/10.1088/1361-648X/aafd6a} {\bibfield  {journal} {\bibinfo
  {journal} {J. Phys. Condens. Matter}\ }\textbf {\bibinfo {volume} {31}},\
  \bibinfo {pages} {163001} (\bibinfo {year} {2019})}\BibitemShut {NoStop}%
\bibitem [{\citenamefont {Oriekhov}\ \emph {et~al.}(2021)\citenamefont
  {Oriekhov}, \citenamefont {Cheipesh},\ and\ \citenamefont
  {Beenakker}}]{oriekhov_voltage_2021}%
  \BibitemOpen
  \bibfield  {author} {\bibinfo {author} {\bibfnamefont {D.~O.}\ \bibnamefont
  {Oriekhov}}, \bibinfo {author} {\bibfnamefont {Y.}~\bibnamefont {Cheipesh}},\
  and\ \bibinfo {author} {\bibfnamefont {C.~W.~J.}\ \bibnamefont {Beenakker}},\
  }\bibfield  {title} {\bibinfo {title} {Voltage staircase in a current-biased
  quantum-dot {Josephson} junction},\ }\href
  {https://doi.org/10.1103/PhysRevB.103.094518} {\bibfield  {journal} {\bibinfo
   {journal} {Phys. Rev. B}\ }\textbf {\bibinfo {volume} {103}},\ \bibinfo
  {pages} {094518} (\bibinfo {year} {2021})}\BibitemShut {NoStop}%
\bibitem [{\citenamefont {Zazunov}\ \emph {et~al.}(2005)\citenamefont
  {Zazunov}, \citenamefont {Shumeiko}, \citenamefont {Wendin},\ and\
  \citenamefont {Bratus’}}]{zazunov_dynamics_2005}%
  \BibitemOpen
  \bibfield  {author} {\bibinfo {author} {\bibfnamefont {A.}~\bibnamefont
  {Zazunov}}, \bibinfo {author} {\bibfnamefont {V.~S.}\ \bibnamefont
  {Shumeiko}}, \bibinfo {author} {\bibfnamefont {G.}~\bibnamefont {Wendin}},\
  and\ \bibinfo {author} {\bibfnamefont {E.~N.}\ \bibnamefont {Bratus’}},\
  }\bibfield  {title} {\bibinfo {title} {Dynamics and phonon-induced
  decoherence of {Andreev} level qubit},\ }\href
  {https://doi.org/10.1103/PhysRevB.71.214505} {\bibfield  {journal} {\bibinfo
  {journal} {Phys. Rev. B}\ }\textbf {\bibinfo {volume} {71}},\ \bibinfo
  {pages} {214505} (\bibinfo {year} {2005})}\BibitemShut {NoStop}%
\bibitem [{\citenamefont {Beenakker}\ and\ \citenamefont {van
  Houten}(1992)}]{beenakker_superconducting_1992}%
  \BibitemOpen
  \bibfield  {author} {\bibinfo {author} {\bibfnamefont {C.}~\bibnamefont
  {Beenakker}}\ and\ \bibinfo {author} {\bibfnamefont {H.}~\bibnamefont {van
  Houten}},\ }\bibfield  {title} {\bibinfo {title} {The {Superconducting}
  {Quantum} {Point} {Contact}},\ }in\ \href
  {https://doi.org/10.1016/B978-0-12-409660-8.50051-1} {\emph {\bibinfo
  {booktitle} {Nanostructures and {Mesoscopic} {Systems}}}}\ (\bibinfo
  {publisher} {Elsevier},\ \bibinfo {year} {1992})\ pp.\ \bibinfo {pages}
  {481--497}\BibitemShut {NoStop}%
\bibitem [{zet()}]{zeta-footnote}%
  \BibitemOpen
  \href@noop {} {\bibinfo {title} {For ${\Gamma}_i \ll
  \sqrt{\Delta^2-\epsilon_g^2}$, $\zeta(\phi)$ depends weakly on the phase:
  $\zeta = \sqrt{\Delta^2-\epsilon_g^2} + {\Gamma} \frac{\epsilon_g^2}{\Delta^2
  - \epsilon_g^2} + {{\cal O}}\left(\frac{\Gamma_i^2}{\Delta^2 - \epsilon_g^2}
  \sin^2\frac{\phi}{2} \right)$}}\BibitemShut {NoStop}%
\bibitem [{\citenamefont {Eckern}\ \emph {et~al.}(2005)\citenamefont {Eckern},
  \citenamefont {Gruber},\ and\ \citenamefont
  {Schwab}}]{eckern_effective_2005}%
  \BibitemOpen
  \bibfield  {author} {\bibinfo {author} {\bibfnamefont {U.}~\bibnamefont
  {Eckern}}, \bibinfo {author} {\bibfnamefont {M.}~\bibnamefont {Gruber}},\
  and\ \bibinfo {author} {\bibfnamefont {P.}~\bibnamefont {Schwab}},\
  }\bibfield  {title} {\bibinfo {title} {Effective descriptions of complex
  quantum systems: path integrals and operator ordering problems},\ }\href
  {https://doi.org/10.1002/andp.200551709-1006} {\bibfield  {journal} {\bibinfo
   {journal} {Annalen der Physik}\ }\textbf {\bibinfo {volume} {517}},\
  \bibinfo {pages} {578} (\bibinfo {year} {2005})}\BibitemShut {NoStop}%
\bibitem [{\citenamefont {Berke}\ \emph {et~al.}(2022)\citenamefont {Berke},
  \citenamefont {Varvelis}, \citenamefont {Trebst}, \citenamefont {Altland},\
  and\ \citenamefont {DiVincenzo}}]{berke2022transmon}%
  \BibitemOpen
  \bibfield  {author} {\bibinfo {author} {\bibfnamefont {C.}~\bibnamefont
  {Berke}}, \bibinfo {author} {\bibfnamefont {E.}~\bibnamefont {Varvelis}},
  \bibinfo {author} {\bibfnamefont {S.}~\bibnamefont {Trebst}}, \bibinfo
  {author} {\bibfnamefont {A.}~\bibnamefont {Altland}},\ and\ \bibinfo {author}
  {\bibfnamefont {D.~P.}\ \bibnamefont {DiVincenzo}},\ }\bibfield  {title}
  {\bibinfo {title} {Transmon platform for quantum computing challenged by
  chaotic fluctuations},\ }\href {https://doi.org/10.1038/s41467-022-29940-y}
  {\bibfield  {journal} {\bibinfo  {journal} {Nat. Commun.}\ }\textbf {\bibinfo
  {volume} {13}},\ \bibinfo {pages} {2495} (\bibinfo {year}
  {2022})}\BibitemShut {NoStop}%
\bibitem [{non()}]{non-perturbative_nz}%
  \BibitemOpen
  \href@noop {} {\bibinfo {title} {The occupation-dependent charge offset $(2
  e) n_z$ is non-perturbative in ${\Gamma}_{L,R}$. $n_z$ originates from the
  $\partial_\tau \phi(\tau)$-term in {Eq}.~\eqref{eq:Gdda}; it can reach $e/2$
  for ${\Gamma}_{L,R} \gg {\Delta}$ and $\delta {\Gamma} \sim
  {\Gamma}$.}}\BibitemShut {Stop}%
\bibitem [{Note2()}]{Note2}%
  \BibitemOpen
  \bibinfo {note} {By projecting onto the odd parity subspace, $\{|\uparrow
  \rangle $, $|\downarrow \rangle \}$, we obtain $\protect \hat H_\protect
  \text {odd}=4 \protect \tilde E_C\left (-i \partial _\phi - \protect \tilde
  n_g \right )^2 + E_\protect \text {cont}(\phi )$, which is relevant when the
  Coulomb interaction is negligible.}\BibitemShut {Stop}%
\bibitem [{Note3()}]{Note3}%
  \BibitemOpen
  \bibinfo {note} {A more rigorous definition of $E_{J,{\protect \rm eff}}$
  with exact continuum contributions does not alter this result.}\BibitemShut
  {Stop}%
\bibitem [{\citenamefont {Fatemi}\ \emph {et~al.}(2025)\citenamefont {Fatemi},
  \citenamefont {Kurilovich}, \citenamefont {Akhmerov},\ and\ \citenamefont
  {van Heck}}]{fatemi_nonlinearity_2024}%
  \BibitemOpen
  \bibfield  {author} {\bibinfo {author} {\bibfnamefont {V.}~\bibnamefont
  {Fatemi}}, \bibinfo {author} {\bibfnamefont {P.~D.}\ \bibnamefont
  {Kurilovich}}, \bibinfo {author} {\bibfnamefont {A.~R.}\ \bibnamefont
  {Akhmerov}},\ and\ \bibinfo {author} {\bibfnamefont {B.}~\bibnamefont {van
  Heck}},\ }\bibfield  {title} {\bibinfo {title} {{Nonlinearity of transparent
  SNS weak links decreases sharply with length}},\ }\href
  {https://doi.org/10.21468/SciPostPhys.18.3.091} {\bibfield  {journal}
  {\bibinfo  {journal} {SciPost Phys.}\ }\textbf {\bibinfo {volume} {18}},\
  \bibinfo {pages} {091} (\bibinfo {year} {2025})}\BibitemShut {NoStop}%
\bibitem [{\citenamefont {Casparis}\ \emph {et~al.}(2018)\citenamefont
  {Casparis}, \citenamefont {Connolly}, \citenamefont {Kjaergaard},
  \citenamefont {Pearson}, \citenamefont {Kringh{\o}j}, \citenamefont {Larsen},
  \citenamefont {Kuemmeth}, \citenamefont {Wang}, \citenamefont {Thomas},
  \citenamefont {Gronin} \emph {et~al.}}]{casparis2018superconducting}%
  \BibitemOpen
  \bibfield  {author} {\bibinfo {author} {\bibfnamefont {L.}~\bibnamefont
  {Casparis}}, \bibinfo {author} {\bibfnamefont {M.~R.}\ \bibnamefont
  {Connolly}}, \bibinfo {author} {\bibfnamefont {M.}~\bibnamefont
  {Kjaergaard}}, \bibinfo {author} {\bibfnamefont {N.~J.}\ \bibnamefont
  {Pearson}}, \bibinfo {author} {\bibfnamefont {A.}~\bibnamefont
  {Kringh{\o}j}}, \bibinfo {author} {\bibfnamefont {T.~W.}\ \bibnamefont
  {Larsen}}, \bibinfo {author} {\bibfnamefont {F.}~\bibnamefont {Kuemmeth}},
  \bibinfo {author} {\bibfnamefont {T.}~\bibnamefont {Wang}}, \bibinfo {author}
  {\bibfnamefont {C.}~\bibnamefont {Thomas}}, \bibinfo {author} {\bibfnamefont
  {S.}~\bibnamefont {Gronin}}, \emph {et~al.},\ }\bibfield  {title} {\bibinfo
  {title} {Superconducting gatemon qubit based on a proximitized
  two-dimensional electron gas},\ }\href
  {https://doi.org/10.1038/s41565-018-0207-y} {\bibfield  {journal} {\bibinfo
  {journal} {Nat. Nanotechnol.}\ }\textbf {\bibinfo {volume} {13}},\ \bibinfo
  {pages} {915} (\bibinfo {year} {2018})}\BibitemShut {NoStop}%
\bibitem [{\citenamefont {Strickland}\ \emph {et~al.}(2024)\citenamefont
  {Strickland}, \citenamefont {Baker}, \citenamefont {Lee}, \citenamefont
  {Dindial}, \citenamefont {Elfeky}, \citenamefont {Strohbeen}, \citenamefont
  {Hatefipour}, \citenamefont {Yu}, \citenamefont {Levy}, \citenamefont
  {Issokson}, \citenamefont {Manucharyan},\ and\ \citenamefont
  {Shabani}}]{PhysRevResearch.6.023094}%
  \BibitemOpen
  \bibfield  {author} {\bibinfo {author} {\bibfnamefont {W.~M.}\ \bibnamefont
  {Strickland}}, \bibinfo {author} {\bibfnamefont {L.~J.}\ \bibnamefont
  {Baker}}, \bibinfo {author} {\bibfnamefont {J.}~\bibnamefont {Lee}}, \bibinfo
  {author} {\bibfnamefont {K.}~\bibnamefont {Dindial}}, \bibinfo {author}
  {\bibfnamefont {B.~H.}\ \bibnamefont {Elfeky}}, \bibinfo {author}
  {\bibfnamefont {P.~J.}\ \bibnamefont {Strohbeen}}, \bibinfo {author}
  {\bibfnamefont {M.}~\bibnamefont {Hatefipour}}, \bibinfo {author}
  {\bibfnamefont {P.}~\bibnamefont {Yu}}, \bibinfo {author} {\bibfnamefont
  {I.}~\bibnamefont {Levy}}, \bibinfo {author} {\bibfnamefont {J.}~\bibnamefont
  {Issokson}}, \bibinfo {author} {\bibfnamefont {V.~E.}\ \bibnamefont
  {Manucharyan}},\ and\ \bibinfo {author} {\bibfnamefont {J.}~\bibnamefont
  {Shabani}},\ }\bibfield  {title} {\bibinfo {title} {Characterizing losses in
  {InAs} two-dimensional electron gas-based gatemon qubits},\ }\href
  {https://doi.org/10.1103/PhysRevResearch.6.023094} {\bibfield  {journal}
  {\bibinfo  {journal} {Phys. Rev. Res.}\ }\textbf {\bibinfo {volume} {6}},\
  \bibinfo {pages} {023094} (\bibinfo {year} {2024})}\BibitemShut {NoStop}%
\bibitem [{\citenamefont {Liu}\ \emph {et~al.}(2025)\citenamefont {Liu},
  \citenamefont {Bordoloi}, \citenamefont {Issokson}, \citenamefont {Levy},
  \citenamefont {Vavilov}, \citenamefont {Shabani},\ and\ \citenamefont
  {Manucharyan}}]{liu2025strongly}%
  \BibitemOpen
  \bibfield  {author} {\bibinfo {author} {\bibfnamefont {S.}~\bibnamefont
  {Liu}}, \bibinfo {author} {\bibfnamefont {A.}~\bibnamefont {Bordoloi}},
  \bibinfo {author} {\bibfnamefont {J.}~\bibnamefont {Issokson}}, \bibinfo
  {author} {\bibfnamefont {I.}~\bibnamefont {Levy}}, \bibinfo {author}
  {\bibfnamefont {M.~G.}\ \bibnamefont {Vavilov}}, \bibinfo {author}
  {\bibfnamefont {J.}~\bibnamefont {Shabani}},\ and\ \bibinfo {author}
  {\bibfnamefont {V.}~\bibnamefont {Manucharyan}},\ }\bibfield  {title}
  {\bibinfo {title} {Strongly-anharmonic gateless gatemon qubits based on
  {InAs/Al 2D} heterostructure},\ }\href {https://arxiv.org/abs/2503.12288}
  {\bibfield  {journal} {\bibinfo  {journal} {arXiv:2503.12288}\ } (\bibinfo
  {year} {2025})}\BibitemShut {NoStop}%
\end{thebibliography}%

\end{document}